\pdfoutput=1
\documentclass[reprint,prb,superscriptaddress,amsmath,amssymb,floatfix]{revtex4-2}
\usepackage[T1]{fontenc}
\usepackage{booktabs,graphicx,placeins,xcolor,needspace}
\usepackage{newtxtext,newtxmath,microtype}
\newcommand{\savefiguremathcodes}[1]{\edef#1{%
 \mathcode`0=\the\mathcode`0\relax\mathcode`1=\the\mathcode`1\relax
 \mathcode`2=\the\mathcode`2\relax\mathcode`3=\the\mathcode`3\relax
 \mathcode`4=\the\mathcode`4\relax\mathcode`5=\the\mathcode`5\relax
 \mathcode`6=\the\mathcode`6\relax\mathcode`7=\the\mathcode`7\relax
 \mathcode`8=\the\mathcode`8\relax\mathcode`9=\the\mathcode`9\relax
 \mathcode`!=\the\mathcode`!\relax\mathcode`,=\the\mathcode`,\relax
 \mathcode`.=\the\mathcode`.\relax\mathcode`:=\the\mathcode`:\relax
 \mathcode`;=\the\mathcode`;\relax\mathcode`?=\the\mathcode`?\relax}}
\savefiguremathcodes{\manuscriptmathcodes}
\usepackage{sansmath}
\savefiguremathcodes{\figuremathcodes}
\manuscriptmathcodes
\usepackage{tikz,pgfplots,algpseudocode}
\usetikzlibrary{arrows.meta,positioning,calc,decorations.markings}
\pgfplotsset{compat=1.18}
\usepackage[colorlinks=true,citecolor=blue!50!black,urlcolor=blue!50!black,linkcolor=blue!50!black]{hyperref}
\newcommand{\Bcal}{\mathcal{B}}
\newcommand{\phbar}{\overline{\mathrm{ph}}}
\newcommand{\ii}{\mathrm{i}}
\newcommand{\rev}[1]{#1}
\newenvironment{revblock}{\par}{\par}
\newcounter{paperalgorithm}
\newenvironment{paperalgorithm}[1]{\par\medskip\noindent\begin{minipage}{\columnwidth}\refstepcounter{paperalgorithm}\textbf{Algorithm \thepaperalgorithm. #1}\par\smallskip\hrule\smallskip\small}{\smallskip\hrule\end{minipage}\par\medskip}
\makeatletter
\newcommand{\algwrap}[1]{\parbox[t]{\dimexpr\linewidth-\ALG@thistlm\relax}{\hangindent=1em\hangafter=1\strut#1\strut}}
\makeatother
\begin{document}

\title{Quantics tensor trains for multiloop functional renormalization group in multiorbital impurity models}

\author{Hyun-Yong Lee}
\email{hyunyong@korea.ac.kr}
\affiliation{Division of Semiconductor Physics, Korea University, Sejong 30019, Korea}
\affiliation{Department of Applied Physics, Graduate School, Korea University, Sejong 30019, Korea}
\affiliation{Center for Extreme Quantum Matter and Functionality, Sungkyunkwan University, Suwon 16419, Korea}

\author{Naoki Kawashima}
\email{kawashima@issp.u-tokyo.ac.jp}
\affiliation{Institute for Solid State Physics, The University of Tokyo, Kashiwa, Chiba 277-8581, Japan}
\affiliation{Trans-scale Quantum Science Institute, The University of Tokyo, Tokyo 113-0033, Japan}

\begin{abstract}
The full frequency dependence of electronic interactions is central to dynamical correlations, but its numerical cost restricts multiorbital calculations. We formulate the multiloop functional renormalization group directly in the quantics tensor train representation, retaining all three independent vertex frequencies over broad ranges without an explicit high-frequency parametrization. Analytical tensor operators perform frequency transformations and internal summations without dense vertex intermediates. Multiloop corrections improve agreement of vertices and responses with independent references and reproduce the low-energy numerical renormalization group self-energy. Lower-temperature calculations resolve a quadratic frequency contribution to the analytically continued scattering rate. A degenerate two-orbital impurity exhibits distinct spin, orbital, and charge responses at fixed Hund coupling. In a compensated altermagnetic environment, an orbital-level splitting induces a local spin response despite zero magnetization of the unperturbed impurity. Increasing Hund coupling suppresses the orbital response while increasing its spin polarization. Renormalized one-particle propagation alone does not reproduce this strong polarization dependence. These results extend compressed multiloop calculations to dynamical correlations in multiorbital environments.
\end{abstract}

\maketitle

\section{Introduction}

\begin{figure*}[!t]
\centering
\begin{minipage}[t]{.675\textwidth}
\vspace{0pt}
\includegraphics[width=\linewidth]{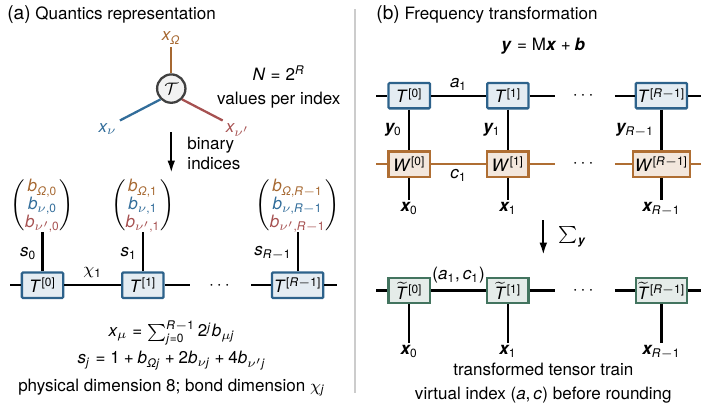}
\end{minipage}\hfill
\begin{minipage}[t]{.315\textwidth}
\vspace{0pt}\centering
\begingroup\fontfamily{qhv}\selectfont\sansmath\figuremathcodes
\mathcode`(="4028 \mathcode`)="5029
\delcode`(="028300 \delcode`)="029301
\begin{tikzpicture}[font=\fontsize{8.54}{10.1}\selectfont]
\begin{semilogyaxis}[width=\linewidth,height=5.9cm,
xmin=6.7,xmax=12.3,ymin=.006,ymax=20000,
xtick={7,8,9,10,12},xlabel={$R$},ylabel={Storage (GiB)},
tick label style={font=\fontsize{8.54}{10.1}\selectfont},
label style={font=\fontsize{8.54}{10.1}\selectfont},
legend style={font=\fontsize{8.54}{10.1}\selectfont,draw=none,at={(.03,.97)},anchor=north west},
title={(c)},title style={at={(0,1)},anchor=south west,font=\fontsize{9.61}{11.2}\selectfont\mdseries},
axis line style={line width=1.05pt},tick style={line width=1.05pt},
grid=major,grid style={gray!18,line width=1.05pt}]
\addplot[black,line width=1.2pt,mark=square*,mark size=1.5pt] coordinates {(7,.1875)(8,1.5)(9,12)(10,96)(12,6144)};
\addlegendentry{Dense vertices}
\addplot[blue!65!black,line width=1.2pt,mark=*,mark size=1.5pt] coordinates {(7,.01614)(8,.02151)(9,.02704)(10,.02750)(12,.02018)};
\addlegendentry{QTT representation}
\end{semilogyaxis}
\end{tikzpicture}
\endgroup
\end{minipage}
\caption{Quantics representation and compressed frequency transformations. (a) Binary digits at the same significance form the local indices of a tensor train. (b) Contraction with the analytical carry matrix product operator changes \rev{the frequency coordinates, i.e., the frequency variables of a channel parametrization; $\boldsymbol x_j$ and $\boldsymbol y_j$ denote the binary digits at site $j$ of the target and source coordinates, and} virtual bonds combine tensor and carry indices. (c) {Storage requirements for the two-loop QTT representation and for six dense vertex amplitudes on grids with $N=2^R$ points per frequency axis}. Dense storage is an estimate, not peak memory or runtime.}
\label{fig:method}
\end{figure*}

Electronic correlations modify both the propagation of individual electrons and the collective response of a material. The self-energy describes the former, whereas the two-particle vertex determines how charge, spin, and orbital fluctuations interact. {Both quantities connect microscopic models to spectroscopy and dynamical response functions\cite{Georges1996,Metzner2012,Georges2013}. Quantum impurity models describe a correlated local degree of freedom coupled to an electronic environment and form the central auxiliary problem of dynamical mean-field theory\cite{Anderson1961,Georges1996,Bulla2008}.} {In multiorbital systems, Hund coupling and orbital-dependent hybridization introduce distinct spin, orbital, and charge dynamics.}

The frequency dependence of the two-particle vertex is a major computational obstacle. Even for a local model, energy conservation leaves three independent frequency arguments in addition to spin and orbital indices. The vertex contains structures associated with different scattering channels, which must be transformed into one another when interchannel feedback is retained. \rev{On a dense grid with $N$ points per frequency axis, each vertex amplitude of each channel requires $N^3$ entries, a change of channel coordinates processes all of them, and each internal Matsubara-frequency sum in a channel contraction requires of order $N^4$ operations. These costs multiply with the number of independent spin and orbital amplitudes, even for models with only a few orbitals.} High-frequency parametrizations and decompositions in terms of bosonic exchange reduce this cost by exploiting known structure\cite{Wentzell2020,Bonetti2022}. {A complementary numerical strategy compresses the joint frequency dependence while retaining all independent frequency arguments\cite{Shinaoka2023,Rohshap2024}.}

The functional renormalization group (fRG) provides a systematic framework for calculating the self-energy and two-particle vertex. It introduces a continuous scale into the propagator and derives coupled flow equations for the self-energy and vertex\cite{Wetterich1993,Salmhofer2001,Metzner2012}. The parquet approximation (PA) self-consistently combines repeated {particle--particle (pp) and particle--hole (ph) scattering} generated from the bare interaction, including feedback between the scattering channels. The multiloop functional renormalization group (mfRG) systematically incorporates the contributions missing at one loop. \rev{Its contributions to the vertex flow are classified by loop order, the number of propagator pairs they contain, and the flow recovers this approximation when contributions of increasing loop order are added until the results converge}\cite{KuglerMfrgPRL2018,KuglerGeneral2018,KuglerParquet2018}. {Calculations for the single-impurity Anderson model with full Matsubara-frequency dependence have demonstrated this convergence at weak to intermediate coupling\cite{ChalupaGantner2022}.} {In grid-based implementations, separating high-frequency asymptotics reduces the cost, but the remaining three-frequency functions still require substantial storage and repeated contractions\cite{Wentzell2020,ChalupaGantner2022}. Extending this explicitly represented frequency domain becomes particularly demanding at low temperatures and with increasing numbers of spin and orbital components.}

Tensor-network representations offer a way to reduce this computational cost by storing frequency-dependent functions compactly and performing operations directly on the compressed tensors\cite{Shinaoka2023,Rohshap2024}. Quantics tensor trains (QTTs) use the binary digits of frequency-grid indices to organize this representation across different scales\cite{Khoromskij2011,Oseledets2011,Shinaoka2023}. Tensor cross interpolation provides an efficient construction for suitable multivariate functions\cite{NunezFernandez2022,Ritter2024}. Low-rank representations and efficient integration of self-energy diagrams have been demonstrated in multiorbital electron--phonon models\cite{Ishida2025}. QTT operations have also been combined into a self-consistent solution of the parquet equations\cite{Rohshap2024}, and numerical-renormalization-group vertices have been compressed by quantics tensor cross interpolation\cite{Frankenbach2025}.

\par
Here we formulate mfRG scale evolution in QTTs, retaining the self-energy, channel-resolved vertices, and their scale derivatives in compressed form. Tensor operators change \rev{the frequency coordinates, that is, the bosonic and fermionic frequency variables that parametrize the vertex in a given scattering channel} [Fig.~\ref{fig:method}(a,b)], while pairwise contractions perform internal Matsubara sums without dense frequency intermediates. {The resulting flows retain the full three-frequency dependence over wide frequency ranges without an explicit high-frequency parametrization of the vertex. The underlying mfRG equations are unchanged, with finite frequency ranges and tensor truncations entering as numerical approximations.}
\par

\begin{revblock}
The aim of this formulation is multiloop calculations for multiorbital impurities, in which the number of independent spin and orbital amplitudes multiplies the cost of every frequency operation. The QTT solution of the parquet equations was demonstrated for the Hubbard atom and the single-impurity Anderson model, and orbital degrees of freedom and the fRG were identified there as future applications\cite{Rohshap2024}. We apply the same compressed frequency operations, without modification, to a degenerate Hund impurity and to an impurity in an altermagnetic environment, where the reduced symmetry leaves 24 spin and orbital amplitudes per channel. The altermagnetic calculations extend to four loops with 1024 points per frequency axis, for which dense storage of the 72 channel amplitudes alone would exceed 1~TiB. The flow also resolves the loop-order and scale dependence of the vertex, self-energy, and responses.
\end{revblock}

\par
We establish the formulation through independent comparisons and then apply the same frequency operations to multiorbital impurities. For the single-impurity Anderson model, PA and numerical renormalization group (NRG) references assess the vertex, dynamical susceptibilities, and self-energy\cite{Ge2024,Ge2023}. Two-loop corrections improve the vertex and responses, while higher loops substantially improve the low-energy self-energy. Lower-temperature calculations resolve a quadratic real-frequency contribution whose coefficient depends on loop order. A degenerate two-orbital impurity exhibits distinct spin, orbital, and charge responses at a fixed Hund coupling. We then examine the spin response to orbital-level splitting in a compensated altermagnetic environment. Directional orbital hybridization retains the spin dependence of the host despite zero magnetization of the unperturbed impurity. Increasing Hund coupling suppresses the orbital response while increasing its spin polarization. The corresponding bubble constructed from the same dressed propagators lacks this strong polarization dependence, demonstrating the role of the two-particle vertex.
\par

\section{Results}
\par
\subsection{Impurity models and fRG flow equations}

We consider Anderson impurities with local interactions coupled to a noninteracting electronic bath,
\begin{equation}
\begin{aligned}
 H={}&\sum_{m\sigma}\epsilon_{dm}n_{m\sigma}+H_{\mathrm{int}}
       +\sum_{\lambda\sigma}\epsilon_{\lambda\sigma}
        c^\dagger_{\lambda\sigma}c_{\lambda\sigma}\\
 &+\sum_{m\lambda\sigma}\left(
 V_{m\lambda\sigma}d^\dagger_{m\sigma}c_{\lambda\sigma}
 +\mathrm{H.c.}\right).
\end{aligned}
\label{main:impurity-hamiltonian}
\end{equation}
Here $d_{m\sigma}$ and $c_{\lambda\sigma}$ annihilate electrons of spin $\sigma$ in impurity orbital $m$ and bath eigenstate $\lambda$, respectively, and $n_{m\sigma}=d^\dagger_{m\sigma}d_{m\sigma}$. Their one-particle energies are $\epsilon_{dm}$ and $\epsilon_{\lambda\sigma}$, while $V_{m\lambda\sigma}$ is the hybridization amplitude. The choices of $H_{\mathrm{int}}$ and bath distinguish the applications below.

The fRG determines how the self-energy $\Sigma_\Lambda$ and four-point vertex $\Gamma_\Lambda$ evolve as a regulator $\Lambda$ is reduced\cite{Metzner2012,Ge2024}. For the flowing self-energy and vertices, a dot denotes differentiation with respect to $\Lambda$. For a regulated inverse bare propagator $Q_{0,\Lambda}$, the dressed propagator is $G_\Lambda=(Q_{0,\Lambda}-\Sigma_\Lambda)^{-1}$ and the single-scale propagator is $S_\Lambda=(\partial_\Lambda G_\Lambda)_\Sigma=-G_\Lambda\dot Q_{0,\Lambda}G_\Lambda$. The first two equations of the exact hierarchy have the structure
\begin{equation}
\begin{aligned}
 \dot\Sigma_\Lambda&=\Gamma_\Lambda\mathbin{\cdot}S_\Lambda,\\
 \dot\Gamma_\Lambda&=\sum_r\Bcal_r[\Gamma_\Lambda,
 \dot\Pi_{r,S,\Lambda},\Gamma_\Lambda]
 +\mathcal T_{6,\Lambda}.
\end{aligned}
\label{main:frg-hierarchy}
\end{equation}
The symbol $\mathbin{\cdot}$ denotes an internal frequency and flavor contraction, where a flavor is the combined spin and orbital index. The bubble $\Bcal_r$ connects two vertices by a propagator pair in the $\mathrm{pp}$, direct $\mathrm{ph}$, or crossed $\mathrm{ph}$ ($\phbar$) channel, including the channel sign and combinatorial factor. Here $\dot\Pi_{r,S,\Lambda}=(\partial_\Lambda\Pi_{r,\Lambda})_\Sigma=S_\Lambda G_\Lambda+G_\Lambda S_\Lambda$, with the frequency assignments of channel $r$, and $\mathcal T_{6,\Lambda}$ is the complete signed contraction of the one-particle-irreducible six-point vertex with $S_\Lambda$. Neglecting this term gives the conventional one-loop truncation; the exact hierarchy also contains flows of higher vertices.

The complete multiloop flow recovers the PA when the loop expansion and self-energy feedback are converged\cite{KuglerGeneral2018,KuglerParquet2018}. Within this approximation the fully irreducible vertex is the bare interaction $\Gamma_0$, and
\begin{equation}
 \Gamma_\Lambda=\Gamma_0+\sum_r\gamma_{r,\Lambda}.
 \label{main:channel-decomposition}
\end{equation}
\nopagebreak
\begin{center}
\begin{tikzpicture}[x=1cm,y=1cm,line width=1.05pt,
  every node/.style={font=\scriptsize,inner sep=1pt},
  subdiagram/.style={draw,line width=1.05pt,fill=blue!10,
    minimum width=3.8mm,minimum height=6.8mm,inner sep=0pt},
  fermion/.style={line width=1.05pt,postaction={decorate},
    decoration={markings,mark=at position .52 with
      {\arrow{Stealth[length=2.8pt,width=2.6pt]}}}},
  cut/.style={densely dashed,line width=1.05pt}]
 \foreach \offset/\channel/\upperleft/\lowerleft/\upperright/\lowerright
   in {0/\mathrm{ph}/1'/1/2'/2,5.0/\phbar/2'/1/1'/2} {
  \begin{scope}[xshift=\offset cm]
   \node[subdiagram] at (.55,0) {};
   \node[subdiagram] at (1.5,0) {};
   \draw[fermion] (.36,.2) -- (0,.46) node[above] {$\upperleft$};
   \draw[fermion] (0,-.46) node[below] {$\lowerleft$} -- (.36,-.2);
   \draw[fermion] (1.69,.2) -- (2.05,.46) node[above] {$\upperright$};
   \draw[fermion] (2.05,-.46) node[below] {$\lowerright$} -- (1.69,-.2);
   \draw[fermion] (.74,.2) -- (1.31,.2);
   \draw[fermion] (1.31,-.2) -- (.74,-.2);
   \draw[cut] (1.025,-.4) -- (1.025,.4);
   \node[font=\footnotesize] at (1.025,-.92) {$\gamma_{\channel,\Lambda}$};
  \end{scope}
 }
 \begin{scope}[xshift=2.5cm]
  \node[subdiagram] at (.55,0) {};
  \node[subdiagram] at (1.5,0) {};
  \draw[fermion] (0,.46) node[above] {$1$} -- (.36,.2);
  \draw[fermion] (0,-.46) node[below] {$2$} -- (.36,-.2);
  \draw[fermion] (1.69,.2) -- (2.05,.46) node[above] {$1'$};
  \draw[fermion] (1.69,-.2) -- (2.05,-.46) node[below] {$2'$};
  \draw[fermion] (.74,.2) -- (1.31,.2);
  \draw[fermion] (.74,-.2) -- (1.31,-.2);
  \draw[cut] (1.025,-.4) -- (1.025,.4);
  \node[font=\footnotesize] at (1.025,-.92) {$\gamma_{\mathrm{pp},\Lambda}$};
 \end{scope}
\end{tikzpicture}
\end{center}
Each $\gamma_{r,\Lambda}$ contains diagrams separated by cutting two propagators in channel $r$. The sketches show these cuts as dashed lines through pairs of dressed propagators. Primed and unprimed labels denote outgoing and incoming external legs, respectively. Shaded blocks represent schematic subdiagrams and do not restrict the perturbative order. This classification is general; model symmetries simplify its spin and orbital structure.

For the single-impurity Anderson model, one orbital has $H_{\mathrm{int}}=Un_\uparrow n_\downarrow$, local level $\epsilon_d=-U/2$, and a spin-independent flat hybridization of width $\Delta$. Spin-rotation symmetry makes the self-energy spin independent and reduces each channel to
\begin{equation}
\begin{aligned}
 \gamma_{r,\Lambda}&=\gamma_{r,D,\Lambda}\sigma_D+\gamma_{r,X,\Lambda}\sigma_X,\\
 \sigma_D&=\delta_{\sigma'_1\sigma_1}\delta_{\sigma'_2\sigma_2},\qquad
 \sigma_X=\delta_{\sigma'_1\sigma_2}\delta_{\sigma'_2\sigma_1}.
\end{aligned}
\label{main:spin-decomposition}
\end{equation}
Here $\gamma_{r,D,\Lambda}$ and $\gamma_{r,X,\Lambda}$ are the frequency-dependent vertex amplitudes associated with the direct and exchange spin contractions $\sigma_D$ and $\sigma_X$, respectively. Each amplitude retains three independent frequency arguments. The labels $D,X$ distinguish spin contractions, whereas $r$ labels the scattering channel. The primed and unprimed spin indices refer to outgoing and incoming legs. Fermionic antisymmetry relates the full vertex at exchanged legs and frequencies, while particle--hole symmetry imposes additional frequency relations\cite{Ge2024}. Neither symmetry closes the flow hierarchy.

We integrate the coupled flows of these amplitudes and the self-energy, retaining contributions of loop order $\ell$ through maximum order $L$:
\begin{equation}
\begin{aligned}
 \dot\gamma_{r,p,\Lambda}&=\sum_{\ell=1}^{L}\dot\gamma_{r,p,\Lambda}^{(\ell)},
                    \qquad p\in\{D,X\},\\
 \dot\Sigma_\Lambda&=\Gamma_\Lambda\mathbin{\cdot}S_\Lambda
              +\dot\Sigma_{\mathrm c,\Lambda}+\dot\Sigma_{\mathrm{fb},\Lambda}.
\end{aligned}
\label{main:integrated-flow}
\end{equation}
The terms $\dot\Sigma_{\mathrm c,\Lambda}$ and $\dot\Sigma_{\mathrm{fb},\Lambda}$ incorporate central vertex contributions and their self-energy feedback, accounting for diagram derivatives missing from the one-loop self-energy flow within the PA. Their definitions and the vertex recursion are given in Methods. Starting from the bare interaction with vanishing channel corrections and Hartree-subtracted self-energy at a large initial regulator, we integrate to the specified endpoint. The six channel amplitudes and self-energy are represented as QTTs throughout this integration.
\par

\subsection{Frequency-dependent flows in the quantics representation}

\par
The numerical task is to evaluate the contractions entering these coupled flow equations without constructing dense frequency arrays. In $\Bcal_r[A,P,B]$, $A$ and $B$ denote the first and second four-point inputs, and $P_r$ is the propagator pair. All inputs are evaluated at the same scale $\Lambda$, whose index is omitted below.

For each channel $r$, we use a bosonic frequency $\Omega$ and two fermionic frequencies $\nu,\nu'$. In the $\mathrm{ph}$ and $\phbar$ channels, $\Omega$ denotes the transferred frequency, a difference between fermionic frequencies. In the $\mathrm{pp}$ channel, $-\Omega$ is the total frequency of the incoming pair. At temperature $T$, their Matsubara values are $\Omega_m=2m\pi T$ and $\nu_n=(2n+1)\pi T$, with integer $m,n$. The external-leg convention is specified in Methods. We denote the two vertex inputs in the frequency coordinates of channel $r$ by $A_r$ and $B_r$, without selecting only the reducible contribution $\gamma_r$. \rev{The internal fermionic Matsubara frequency $\nu_k$, with integer $k$, is summed over. All frequency sums run over Matsubara indices, and $\omega$ denotes a real frequency throughout.}
\par

\par
For the $\mathrm{ph}$ channel, the propagator pair is \rev{$\Pi_{\mathrm{ph}}(\Omega,\nu_k)=G(\nu_k+\Omega)G(\nu_k)$}, with dressed propagator $G$. At temperature $T$, the contraction is represented by
\begin{equation}
\begin{aligned}
 &\Bcal_{\mathrm{ph}}[A,\Pi,B](\Omega,\nu,\nu')\\[2pt]
 &\quad=T\rev{\sum_k}\quad
 \begin{tikzpicture}[baseline=-2pt,x=1cm,y=1cm,line width=1.05pt,
   every node/.style={font=\footnotesize,inner sep=1pt},
   fermion/.style={line width=1.05pt,postaction={decorate},
     decoration={markings,mark=at position .55 with
       {\arrow{Stealth[length=3.2pt,width=3pt]}}}}]
  \node[draw,minimum width=6mm,minimum height=8.5mm,inner sep=0pt]
    (leftvertex) at (0,0) {$B$};
  \node[draw,minimum width=6mm,minimum height=8.5mm,inner sep=0pt]
    (rightvertex) at (2,0) {$A$};
  \draw[fermion] (.3,.25) -- node[above=2pt] {\rev{$G(\nu_k)$}} (1.7,.25);
  \draw[fermion] (1.7,-.25) -- node[below=2pt] {\rev{$G(\nu_k+\Omega)$}} (.3,-.25);
  \draw[fermion] (-.3,.25) -- (-1.05,.7) node[above] {$\nu+\Omega$};
  \draw[fermion] (-1.05,-.7) node[below] {$\nu$} -- (-.3,-.25);
  \draw[fermion] (2.3,.25) -- (3.05,.7) node[above] {$\nu'$};
  \draw[fermion] (3.05,-.7) node[below] {$\nu'+\Omega$} -- (2.3,-.25);
 \end{tikzpicture}
\end{aligned}
\label{main:bubble-diagram}
\end{equation}
The rectangles denote the four-point inputs; only the two internal lines carry propagators. The external legs are amputated, and arrows indicate fermion flow. In channel coordinates, the general contraction reads
\begin{equation}
 \begin{aligned}
 &\Bcal_r[A,P,B](\Omega,\nu,\nu') \\
 &= c_r T\rev{\sum_{k} A_r(\Omega,\nu_k,\nu')
 P_r(\Omega,\nu_k)B_r(\Omega,\nu,\tilde\nu_r)}.
 \end{aligned}
\label{main:bubble}
\end{equation}
Here \rev{$\tilde\nu_r=\nu_k$ for the $\mathrm{ph}$ and $\phbar$ channels and $\tilde\nu_{\mathrm{pp}}=-\Omega-\nu_k$, so that the second input is evaluated at the internal frequency of the corresponding channel.} The factor $c_r$ contains the channel sign and combinatorial factor, with $c_{\mathrm{ph}}=1$ in our convention. Internal spin and orbital indices are summed; their contractions are specified in Methods. The one-loop contribution in Eq.~\eqref{main:integrated-flow} is $\dot\gamma_r^{(1)}=\Bcal_r[\Gamma,\dot\Pi_r,\Gamma]$. We include the Katanin substitution: $\dot\Pi_r=\dot G_KG+G\dot G_K$, where $\dot G_K=S+G(\Gamma\mathbin{\cdot}S)G$\cite{Katanin2004}. Higher-loop contributions insert channel-resolved vertex derivatives and use $P_r=\Pi_r=GG$ without a scale derivative. Thus, the same frequency transformations and contractions serve the entire loop hierarchy.
\par

\par
For $N=2^R$ points per frequency axis, we write each index in binary form and group the three bits of the same significance into a local index $s_j$, with $j=0,\ldots,R-1$ ordered from least to most significant [Fig.~\ref{fig:method}(a)]. A channel amplitude, denoted here by $\mathcal T$, becomes
\begin{equation}
 \mathcal T(s_0,\ldots,s_{R-1})
 =\sum_{\alpha_1,\ldots,\alpha_{R-1}}
 T^{[0]}_{1s_0\alpha_1}\cdots
 T^{[R-1]}_{\alpha_{R-1}s_{R-1}1}.
 \label{main:qtt}
\end{equation}
The auxiliary indices $\alpha_j$ have bond dimensions $\chi_j$, with $\chi_0=\chi_R=1$ at the boundaries. Equation~\eqref{main:qtt} is a tensor train, equivalently a matrix product state (MPS), with local physical dimension eight. \rev{For bond dimensions bounded by $\chi$, its storage is at most of order $R\chi^2$}, instead of $N^3$. \rev{Because the bond dimensions themselves depend on the grid, the actual storage need not grow linearly with $R$.} The bond dimensions are numerical quantities determined by the required accuracy.
\par

\par
Changing channel coordinates requires frequency sums and differences, such as $\nu\mapsto\nu+\Omega$ and $\nu\mapsto-\Omega-\nu$. Matrix product operators (MPOs) implement these transformations exactly, with auxiliary indices that transmit arithmetic carries between binary digits [Fig.~\ref{fig:method}(b)]. Subsequent compression, called tensor rounding, reduces the bond dimensions and can introduce truncation errors. In the resulting four-frequency MPS, the pairwise contraction first combines one vertex with the propagator pair, then the second vertex, summing the internal bit at each site. Intermediate bond dimensions and discarded weights are controlled separately from the stored vertex. Methods gives the local tensors and numerical factorization.
\par

{Compression makes a broad three-frequency domain accessible throughout the interacting flow.} Figure~\ref{fig:method}(c) compares {the storage of two-loop QTT representations} over increasing frequency ranges. At $R=12$, the six dense complex vertex amplitudes would require $6$~TiB, whereas the stored {QTT representation}, including the self-energy, occupies about $21$~MiB. \rev{Over $R=7,\ldots,12$, the stored two-loop representation remains between 16.5 and 28.2~MiB and does not grow linearly with $R$. Its entry count $\sum_j8\chi_j\chi_{j+1}$ depends on bond dimensions determined by the truncation tolerance and the frequency structure resolved on each grid; the largest endpoint bond dimension of the two-loop vertex is 157 at $R=8$ and 138 at $R=10$.} {The comparison concerns storage}; temporary arrays in contractions require additional memory. {For the single-impurity Anderson model at $T/U=10^{-2}$, the $R=12$ representation contains 4096 points along each frequency axis, with positive fermionic frequencies extending from $\pi T$ to approximately $129U$.} Because the Matsubara spacing is fixed by temperature, increasing $R$ extends the frequency range rather than refining its spacing. {The full three-frequency dependence is thus retained across the chosen domain without an explicit high-frequency parametrization.}

\subsection[Single-impurity Anderson model]{{Single-impurity Anderson model}}

\par
The particle--hole-symmetric single-impurity Anderson model provides
independent PA and NRG references. We set
$H_{\mathrm{int}}=Un_\uparrow n_\downarrow$ and $\epsilon_d=-U/2$
in Eq.~\eqref{main:impurity-hamiltonian}. The additive regulator increases the
flat hybridization width from $\Delta$ to
$\Delta_{\mathrm{eff}}=\Delta+\Lambda$, so the dimensionless interaction
$u=U/(\pi\Delta_{\mathrm{eff}})$ increases as the regulator is lowered at
fixed $U$. We compare observables near $u=1$ and $T/U=10^{-2}$ at a finite
endpoint $\Lambda\simeq\Delta/2$, with the effective hybridization matched
to the reference calculation\cite{Ge2024,Ge2023}. For the flat bath, the
finite regulator thus specifies the impurity hybridization used in the
comparison.
\par

\begin{figure}[t]
\centering
\includegraphics[width=\columnwidth]{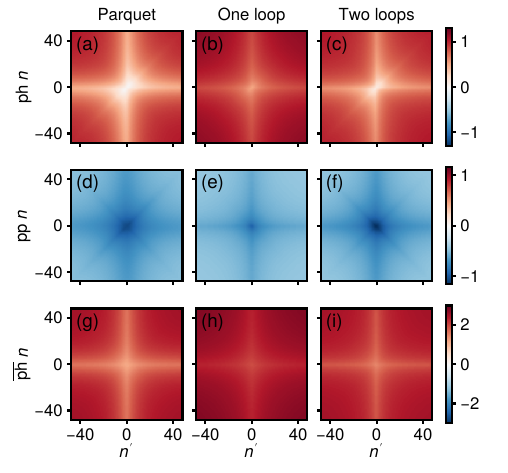}
\caption{Channel-resolved vertex near $u=1$, $R=8$ \rev{($N=256$)}, and $T/U=10^{-2}$. Columns show the parquet reference, one-loop, and two-loop results; {rows show the $\mathrm{ph}$, $\mathrm{pp}$, and $\phbar$ channels}. Colors give {$\operatorname{Re}\gamma_{r,D}/U$} at zero bosonic transfer. {The axes are fermionic Matsubara indices; central frequency windows are displayed.}}
\label{fig:parquet}
\end{figure}

The first multiloop correction improves the vertex over complete
fermionic-frequency planes, beyond selected low-frequency points.
Figure~\ref{fig:parquet} resolves the direct spin-contraction amplitude
in all three channels. {The $\mathrm{pp}$ vertex
develops a pronounced diagonal frequency structure at two loops,
whereas the one-loop result is dominated by horizontal and vertical
features. The combined deviation from the PA reference decreases from
about $16\%$ to $4\%$ at $R=8$ \rev{($N=2^8=256$ points per frequency axis)} upon including the two-loop correction.}
The combined comparison includes every point of
all three channel planes, with no separate channel rescaling.
{Supplementary Fig.~S1 also gives the individual channel deviations.}
Higher loops need not improve
every observable monotonically at fixed tensor dimensions and frequency
range\cite{ChalupaGantner2022}.

\begin{figure*}[t]
\centering
\includegraphics[width=\textwidth]{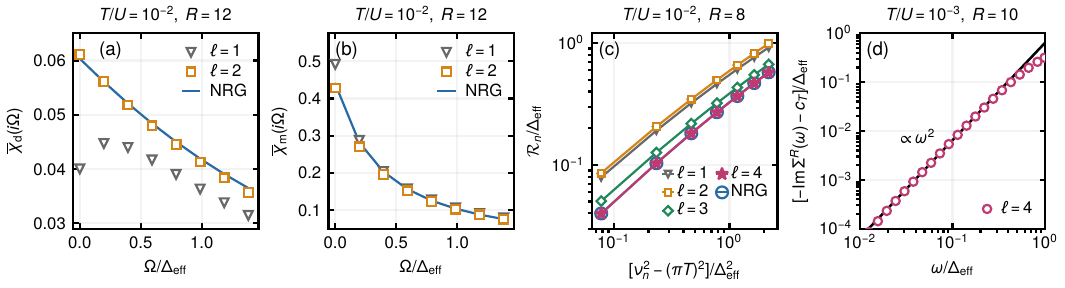}
\caption{Single-impurity dynamics. {(a,b) Charge (d) and magnetic (m) susceptibilities at $R=12$ \rev{($N=4096$) as functions of the bosonic Matsubara frequency $\Omega$}, in the NRG normalization defined in Methods. (c) Self-energy residual $\mathcal R_n$ defined in Eq.~\eqref{main:sigma-residual}, at $R=8$ \rev{($N=256$), plotted against the fermionic Matsubara frequency $\nu_n$}; the four-loop result (stars) reproduces the NRG residual (open circles).} Panels (a--c) use $T/U=10^{-2}$. {(d) Four-loop \rev{scattering rate as a function of real frequency $\omega$} at $T/U=10^{-3}$ and $R=10$ \rev{($N=1024$)}, after Pad\'e continuation and subtraction of the constant $c_T$ (open circles); the black line indicates quadratic dependence. Colors identify loop orders in (a--c).}}
\label{fig:siam-dynamics}
\end{figure*}

{The two-loop correction restores the low-frequency behavior
of the dynamical charge susceptibility, $\overline{\chi}_{\mathrm d}(\ii\Omega)$
[Fig.~\ref{fig:siam-dynamics}(a)]. The one-loop result has a spurious maximum
at the first nonzero Matsubara frequency; two loops recover the monotonic
decrease found in NRG. The magnetic susceptibility,
$\overline{\chi}_{\mathrm m}(\ii\Omega)$, also approaches NRG more closely,
particularly at zero frequency [Fig.~\ref{fig:siam-dynamics}(b)]. Both
responses include vertex contributions summed over two fermionic
frequencies, with normalization defined in Methods.}

\par
Higher-loop corrections substantially improve the low-frequency
self-energy. To resolve its nonlinear frequency dependence, we remove
the linear term fixed by the first Matsubara frequency and define
\begin{equation}
 \mathcal R_n=\operatorname{Im}\Sigma_{\mathrm{odd}}(\ii\nu_n)
 -\frac{\nu_n}{\nu_0}\operatorname{Im}\Sigma_{\mathrm{odd}}(\ii\nu_0),
\label{main:sigma-residual}
\end{equation}
where $\Sigma_{\mathrm{odd}}(\ii\nu)
=[\Sigma(\ii\nu)-\Sigma(-\ii\nu)]/2$ and $\nu_0=\pi T$.
Figure~\ref{fig:siam-dynamics}(c) compares this residual at
$T/U=10^{-2}$ and $R=8$. The four-loop result reproduces the NRG residual,
whereas lower orders remain visibly separated. Each curve uses its own
lowest-frequency value for the subtraction, without fitting its
amplitude to NRG.
\par

\par
At lower temperature, the real-frequency self-energy exhibits the
quadratic frequency dependence characteristic of a Fermi
liquid\cite{Yamada1975a,Hewson1993,Hewson2001}.
Figure~\ref{fig:siam-dynamics}(d) shows the four-loop result at
$T/U=10^{-3}$ and $R=10$, obtained by Pad\'e analytic
continuation\cite{Vidberg1977}. After subtraction of a low-frequency
constant $c_T$, the retarded self-energy follows
$-\operatorname{Im}\Sigma^{\mathrm{ret}}(\omega)-c_T\propto\omega^2$
over the resolved low-frequency interval. The subtraction and
continuation are specified in Methods.
The quadratic frequency dependence is already present at lower loop
orders, but its coefficient remains loop dependent. The NRG comparison
at $T/U=10^{-2}$ [Fig.~\ref{fig:siam-dynamics}(c)] provides a separate
quantitative assessment, showing improved agreement in the nonlinear
Matsubara-frequency dependence at four loops.
\par

\Needspace{6\baselineskip}
\subsection{Contrasting spin and orbital responses in a Hund impurity}

\par
The frequency operations extend to multiple orbitals by treating the spin and orbital sums separately from the frequency dependence. We first consider two degenerate orbitals in Eq.~\eqref{main:impurity-hamiltonian}, with identical spin-independent flat baths and no interorbital hybridization. Their local interaction is the rotationally invariant Kanamori interaction\cite{Kanamori1963,Georges2013},
\begin{align}
 H_{\mathrm{int}}={}&U\sum_a n_{a\uparrow}n_{a\downarrow}
 +(U-2J)\sum_{a\ne b}n_{a\uparrow}n_{b\downarrow}\notag\\
 &+(U-3J)\sum_{a<b,\sigma}n_{a\sigma}n_{b\sigma}
 \label{meth:kanamori}\\
 &-J\sum_{a\ne b}d^\dagger_{a\uparrow}d_{a\downarrow}
 d^\dagger_{b\downarrow}d_{b\uparrow}
 +J\sum_{a\ne b}d^\dagger_{a\uparrow}d^\dagger_{a\downarrow}
 d_{b\downarrow}d_{b\uparrow},\nonumber
\end{align}
where $a,b\in\{1,2\}$, $U$ is the intraorbital repulsion, and $J$ is Hund coupling. The last two terms describe spin flip and pair hopping. The level $\epsilon_d=-(3U-5J)/2$ ensures particle--hole symmetry. Spin rotation and orbital symmetry permit a basis of six tensors in spin and orbital space per scattering channel, specified in Methods. We apply the same QTT transformations and internal Matsubara-frequency sums to the frequency-dependent vertex amplitudes. Fixed coefficients account for the internal spin and orbital sums. No single-frequency approximation is introduced when the orbital degrees of freedom are added.
\par

\par
Hund coupling affects spin, orbital, and charge correlations differently\cite{Georges2013}. We extract the spin, orbital-quadrupole, and charge components of the full vertex by forming linear combinations of its amplitudes, with weights specified in Methods. These components refer to spin and orbital indices, whereas the $\mathrm{pp}$, $\mathrm{ph}$, and $\phbar$ channels organize the vertex frequency dependence. We denote their real parts at zero transfer and the fermionic-frequency pair $(\pi T,-\pi T)$ in $\mathrm{ph}$ coordinates by $g_{\mathrm{sp}}$, $g_{\mathrm{orb}}$, and $g_{\mathrm{ch}}$, respectively. Their bare values are
\begin{equation}
 \begin{aligned}
 g_{\mathrm{ch}}=3U-5J,\,\,\, g_{\mathrm{sp}}=-(U+J),\,\,\, g_{\mathrm{orb}}=-(U-5J).
 \end{aligned}
 \label{main:bare-couplings}
\end{equation}
We use the same flow coordinate $u=U/[\pi(\Delta+\Lambda)]$ as for the single-orbital model, with $U$ denoting the intraorbital interaction.
\par

\begin{figure}[t]
\centering
\includegraphics[width=\columnwidth]{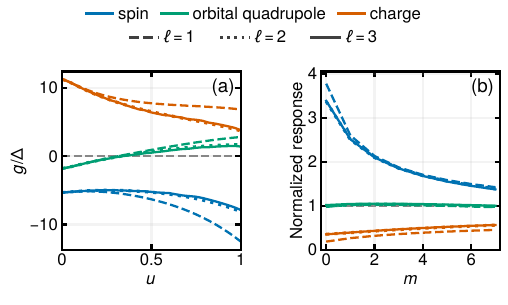}
\caption{Two-orbital Hund impurity at $J/U=0.125$ and $R=8$ \rev{($N=256$)}. {(a) Spin, orbital-quadrupole, and charge interaction components along the flow; $u$ increases as the regulator decreases at fixed $U$.} (b) Dynamical spin, orbital-quadrupole, and charge responses near $u=1$, normalized by their corresponding dressed bubble contributions; $m$ is the bosonic Matsubara index. Line styles distinguish one, two, and three loops. The orbital interaction changes sign along the flow, while the responses retain the hierarchy spin, orbital quadrupole, and charge.}
\label{fig:hund}
\end{figure}

Figure~\ref{fig:hund}(a) shows that the renormalized orbital-quadrupole interaction \rev{$g_{\mathrm{orb}}$, defined above Eq.~\eqref{main:bare-couplings} and in Methods, changes sign between $u=0.31$ and $0.34$ at one to three loops}, while the spin and charge components remain well separated. \rev{Its bare value $-(U-5J)$ is attractive and smaller in magnitude than the bare spin and charge components at $J/U=0.125$. The flow overturns it, so that the renormalized orbital interaction no longer favors orbital fluctuations.} {The corresponding dynamical responses in Fig.~\ref{fig:hund}(b) display a strong spin enhancement, a much weaker orbital response, and suppressed charge fluctuations.} These responses are divided by the vertex-free bubbles constructed from the dressed propagators of the same calculation. This normalization isolates the effect of the vertex relative to the interacting one-particle propagation; it is not a comparison with a noninteracting impurity. \rev{Relative to these bubbles, the vertex keeps the orbital response within 5\% of unity at one to three loops, while it enhances the spin response and suppresses the charge response. The hierarchy of the responses is therefore produced by the vertex renormalization rather than by the bare couplings, which alone would favor orbital fluctuations.}

\par
At fixed $J/U=0.125$ and $R=8$, the normalized dynamical responses show both qualitative and quantitative convergence with respect to loop order by three loops. The two- and three-loop curves nearly coincide over the displayed frequency range, preserving the distinct spin, orbital-quadrupole, and charge responses. This hierarchy is consistent with the established distinction between spin and orbital fluctuations in Hund-coupled impurities\cite{deMedici2011,Georges2013,Stadler2015}. 
\par

\subsection{Spin-selective orbital response in an altermagnetic environment}
\label{am:results-hund}

\par
\begin{figure}[t]
 \centering
 \includegraphics[width=\columnwidth]{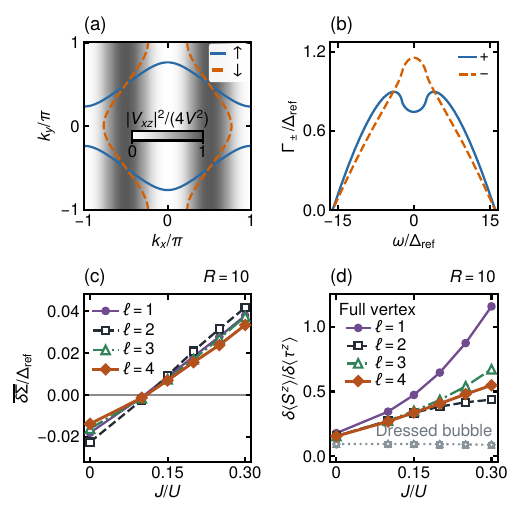}
 \caption{Electronic correlations and orbital response in an altermagnetic environment. (a) Host Fermi contours and $xz$ hybridization weight. {(b) Hybridization widths $\Gamma_\pm=-\operatorname{Im}\Delta_\pm(\omega+i0^+)$ of the two spin--orbital groups\rev{ as functions of real frequency $\omega$}. (c) \rev{Self-energy difference averaged over the Matsubara frequencies $\nu_n$, $n=2,\ldots,8$}. (d) Spin polarization of the orbital-field response. Panels (c,d) show one- to four-loop results at $R=10$ \rev{($N=1024$)}, $T/U=10^{-2}$, and $\Lambda\simeq0.5\Delta_{\mathrm{ref}}$ for six Hund couplings. In (d), the full vertex produces increasing spin selectivity at every loop order, whereas the four-loop dressed bubble (gray) varies weakly with Hund coupling.}}
 \label{fig:am-combined}
\end{figure}

We use directional impurity orbitals to retain the spin dependence of a compensated altermagnetic bath. Altermagnets combine compensated magnetization with momentum-dependent spin splitting\cite{Smejkal2022PRX,Krempasky2024}, and their orbital dependence permits spin-selective responses\cite{Vila2025}. We investigate how electronic correlations modify the local spin response to an orbital-level splitting. A momentum-independent impurity coupling can average away the host spin dependence\cite{Diniz2024}. We instead retain the local interaction of Eq.~\eqref{meth:kanamori} and its particle--hole-symmetric level, but couple an $xz,yz$ doublet to a square-lattice host without spin--orbit coupling. In Eq.~\eqref{main:impurity-hamiltonian} we set $\lambda=\mathbf k$ and $V_{m\mathbf k\sigma}=V_{m,\mathbf k}$. The bath dispersion and spin-independent hybridization amplitudes are
\begin{equation}
\begin{aligned}
 &\epsilon_{\mathbf k\sigma}=-2t(\cos k_x+\cos k_y)+\sigma t_J(\cos k_x-\cos k_y),\\
 &(V_{xz,\mathbf k},V_{yz,\mathbf k})=2\ii V(\sin k_x,\sin k_y).
\end{aligned}
\label{am:host-hamiltonian}
\end{equation}
The lattice constant is unity and $\sigma=\pm1$. The hopping $t$ sets the host energy scale, $t_J$ is the prescribed spin-splitting amplitude, and $V$ sets the impurity--bath coupling. In particular, $t_J$ is distinct from the local Hund coupling $J$. The two hybridization amplitudes select different momentum regions [Fig.~\ref{fig:am-combined}(a,b)], so the impurity retains the host spin dependence through its orbital couplings. \rev{Both spin bands have the same density of states, so $\Gamma_+$ and $\Gamma_-$ in Fig.~\ref{fig:am-combined}(b) differ only through the weight $\sin^2k_x$ averaged over constant-energy contours. At $\omega=0$ the up-spin contour samples a smaller average weight than the down-spin contour. Away from $\omega=0$, $\Gamma_+$ first increases with the density of states toward its van Hove singularity at $|\omega|=2t_J$; beyond it, the contour-averaged weight increases while the density of states decreases, which places the two maxima of $\Gamma_+$ near $|\omega|\approx4\Delta_{\mathrm{ref}}$.} Combined orbital interchange and spin reversal leave two groups, $+=(xz\uparrow,yz\downarrow)$ and $-=(xz\downarrow,yz\uparrow)$, with distinct hybridizations. Particle--hole symmetry fixes half occupation in every flavor, so the unperturbed impurity has no magnetization. The two groups nevertheless have distinct self-energies and can respond differently to an orbital perturbation.
\par

\par
The same compressed frequency operations accommodate this reduced flavor symmetry. Each channel has 24 amplitudes, giving 72 frequency-dependent vertex functions and two self-energies. \rev{These comprise the six tensors of the degenerate model and 18 of their projections onto the two spin--orbital groups, which the group-dependent propagators generate; the 24 tensors span the smallest space containing the bare vertex that is closed under the bubble contractions (Methods), not a product of six tensors and four sign classes.} We vary Hund coupling at fixed $t_J/t=0.3$, $U=1.5\pi\Delta_{\mathrm{ref}}$, and $T/U=10^{-2}$, and evaluate the observables at the common finite regulator endpoint $\Lambda\simeq0.5\Delta_{\mathrm{ref}}$. Here $\Delta_{\mathrm{ref}}$ is the zero-energy hybridization without host splitting.
\par

\par
The self-energy asymmetry between the two spin--orbital groups reverses with Hund coupling at fixed altermagnetic bath. Figure~\ref{fig:am-combined}(c) shows $\overline{\delta\Sigma}/\Delta_{\mathrm{ref}}$, where $\overline{\delta\Sigma}$ is the mean of $\operatorname{Im}[\Sigma_+(i\nu_n)-\Sigma_-(i\nu_n)]$ over $n=2,\ldots,8$\rev{, a window that excludes the lowest Matsubara frequencies (Methods)}. The sign reversal occurs near $J/U=0.11$. Its position lies between $J/U=0.10$ and $0.15$ at one to four loops for $R=10$ and in the one-loop calculations at $R=7$ and $8$, whereas the magnitude of $\overline{\delta\Sigma}$ still changes between three and four loops (Supplementary Note~8). Thus electronic correlations change the self-energy contrast without changing the host spin splitting. \rev{Because $\Gamma_-$ exceeds $\Gamma_+$ at low frequencies [Fig.~\ref{fig:am-combined}(b)], the hybridization gives $\operatorname{Im}[\Delta_+(i\nu_n)-\Delta_-(i\nu_n)]>0$ in the same window. A negative $\overline{\delta\Sigma}$, found for weak Hund coupling, therefore partially compensates this asymmetry in the inverse propagators $G_\pm^{-1}$, whereas a positive $\overline{\delta\Sigma}$ adds to it.} Competing Kanamori scattering processes permit a reversal already at second order. The derivation and the supporting frequency dependence are given in Supplementary Note~4 and Supplementary Fig.~S2 (Supplementary Note~5), respectively.
\par

An orbital-level splitting induces a local spin response despite zero magnetization of the unperturbed impurity. Consider $H'=-h_{\mathrm{orb}}\tau^z$, where $\tau^z=(n_{xz}-n_{yz})/2$, and define $S^z=(n_\uparrow-n_\downarrow)/2$. {Here $n_m=\sum_\sigma n_{m\sigma}$ and $n_\sigma=\sum_m n_{m\sigma}$.} A positive $h_{\mathrm{orb}}$ lowers the $xz$ level and raises the $yz$ level equally for both spins. This perturbation therefore changes the orbital energies without applying a magnetic field. In linear response, the ratio of the induced changes is
\begin{equation}
 \frac{\delta\langle S^z\rangle}{\delta\langle\tau^z\rangle}
 =\frac{\chi_{S^z,\tau^z}(0)}{\chi_{\tau^z,\tau^z}(0)}.
 \label{am:orbital-polarization}
\end{equation}
Symmetry makes this ratio equal to the difference divided by the sum of the spin-resolved $xz$ occupancy changes. Both changes have the same sign at two to four loops, so the ratio measures their spin polarization; at one loop the $xz\downarrow$ change reverses sign at $J/U=0.30$, where the ratio exceeds one. It describes an induced response, not a magnetization of the unperturbed impurity.

\par
The orbital response in the altermagnetic environment decreases in magnitude but becomes increasingly spin polarized with Hund coupling [Fig.~\ref{fig:am-combined}(d)]: at four loops the polarization rises from $0.15$ at $J=0$ to $0.55$ at $J/U=0.30$, where the two- and three-loop values are $0.44$ and $0.67$ (Supplementary Note~8). At a fixed weak orbital field, both $xz$ occupancy responses decrease, but the down-spin response is suppressed more rapidly. The occupancy change therefore becomes more strongly polarized even as the total orbital response decreases. The mixed spin--orbital susceptibility also decreases, but more slowly than the orbital susceptibility, so their ratio increases while the absolute induced magnetization decreases. Supplementary Fig.~S3 resolves these component responses (Supplementary Note~6).

The strong dependence of this spin polarization on Hund coupling requires the two-particle vertex. The dressed bubble constructed from the same propagators has only a weak Hund dependence, remaining near $0.09$ [Fig.~\ref{fig:am-combined}(d)]. Both responses retain the same self-energy corrections, so renormalized one-particle propagation alone does not reproduce the change in spin selectivity. At each calculated Hund coupling, the vertex also enhances the absolute mixed susceptibility relative to its dressed bubble [Supplementary Fig.~S3(b)]. This enhancement is distinct from the decrease of that susceptibility as Hund coupling grows. The loop-order dependence of Fig.~\ref{fig:am-combined}(c,d) is quantified in Supplementary Note~8, and the regulator-cutoff comparison is provided in Supplementary Fig.~S4 (Supplementary Note~7).
\par

\section{Discussion}

\par
{The compressed formulation extends the frequency domain treated explicitly in multiloop flows while retaining the full three-frequency vertex in different orbital environments.} It separates the numerical tensor representation from the physical loop truncation without removing an independent frequency argument.

Independent comparisons demonstrate the importance of multiloop corrections for the vertex, dynamical responses, and low-energy self-energy. In particular, these corrections are essential for reproducing the NRG self-energy beyond its leading Matsubara slope \rev{[Fig.~\ref{fig:siam-dynamics}(c)]}. At lower temperature, the coefficient of the quadratic real-frequency dependence remains sensitive to multiloop corrections, showing that the expected power law alone does not establish quantitative accuracy.
\par

\begin{revblock}
The agreement with NRG in Fig.~\ref{fig:siam-dynamics} must be interpreted relative to the PA. Upon convergence in loop order, mfRG approaches the PA, not the exact solution. For this model, the cited real-frequency study found that the PA reproduces NRG closely for static one- and two-particle properties for $u\lesssim1$, the range in which parquet solutions were obtained. At $u=1$, however, the low-frequency opposite-spin vertex already differs from NRG, and the maxima of the real-frequency PA self-energy are displaced outward, a deviation hardly visible on the Matsubara axis\cite{Ge2024}. The agreement in Fig.~\ref{fig:siam-dynamics} therefore reflects the accuracy of the PA at this coupling and need not persist at stronger coupling. For the multiorbital impurities no independent reference is available here; their accuracy is assessed only through the loop-order dependence and, where stated, the dependence on bond dimension, frequency range, and regulator endpoint.
\end{revblock}

\par
The multiorbital applications establish complementary aspects of the formulation. At a fixed Hund coupling, the degenerate impurity exhibits distinct dynamical spin, orbital-quadrupole, and charge responses through three loop orders \rev{[Fig.~\ref{fig:hund}(b)]}. The altermagnetic application addresses how electronic correlations modify an orbital response that carries spin information despite zero magnetization of the unperturbed impurity.
\par

The spin selectivity of the orbital response in an altermagnetic environment is not determined by its spin-split one-particle structure alone. With the bath held fixed, Hund coupling reverses the self-energy contrast between the spin--orbital groups \rev{[Fig.~\ref{fig:am-combined}(c)]} and increases the polarization of the orbital-field response \rev{[Fig.~\ref{fig:am-combined}(d)]}. The latter results from unequal suppression of the spin-resolved occupancy changes: the orbital response and the absolute induced magnetization decrease, while the response becomes more spin selective. Its strong Hund dependence is absent from the bubble formed with the same dressed propagators, demonstrating the importance of the two-particle vertex.

Orbital perturbations provide a way to probe spin-dependent electronic correlations in a compensated altermagnetic environment. The coupling between orbital and spin degrees of freedom already motivates spin-selective optical probes of altermagnets\cite{Vila2025}, and stress-induced magnetization has been observed in MnTe\cite{Aoyama2024}. The present local calculation identifies how electronic correlations modify the spin composition of the response to an orbital-level splitting. Quantitative predictions for a strain or optical experiment require the corresponding coupling to the orbitals and the host response, beyond the static impurity susceptibility calculated here.

Finite loop order, frequency range, and tensor dimensions remain separate approximations, with analytic continuation entering only the real-frequency \rev{self-energy in Fig.~\ref{fig:siam-dynamics}(d)}. The altermagnetic response extends the multiloop calculation to an orbital-dependent, spin-split bath. Extending the formulation to a self-consistent materials calculation requires the appropriate bath update and an assessment of the interaction regime. Within the impurity setting, the present results already combine a systematic diagrammatic hierarchy with orbital-dependent environments and their one- and two-particle dynamics.

\Needspace{6\baselineskip}
\section{Methods}
\subsection{Models, frequency conventions, and flavor algebra}
\label{meth:models}

\par
We use units $k_{\mathrm B}=\hbar=1$. \rev{At particle--hole symmetry, the static Hartree self-energy cancels the local level, $\epsilon_d+\Sigma_{\mathrm H}=0$, with $\Sigma_{\mathrm H}=U/2$ for one orbital and $(3U-5J)/2$ for two orbitals. Absorbing $\Sigma_{\mathrm H}$ in the local level therefore removes the level from the inverse propagator, which contains $\ii\nu$ alone before the bath is included, and $\Sigma$ denotes the Hartree-subtracted self-energy. Integrating out} the flat bath and adding a hybridization regulator gives
\begin{equation}
 Q_{0,\Lambda}(\ii\nu)=\ii\nu+\ii(\Delta+\Lambda)\operatorname{sgn}\nu.
 \label{meth:regulator}
\end{equation}
For the degenerate two-orbital model this inverse propagator is diagonal
and identical in spin and orbital indices. The corresponding
altermagnetic propagators are specified below.
\par

\par
For the decomposition in Eq.~\eqref{main:channel-decomposition}, writing
outgoing and incoming legs as $(o_1,o_2;i_1,i_2)$ defines the channel
coordinates and internal propagator frequencies \rev{$(\bar\nu_1,\bar\nu_2)$}:
\begin{equation}
\setlength{\arraycolsep}{2pt}
\begin{array}{c|c|c}
r&(o_1,o_2;i_1,i_2)&\rev{(\bar\nu_1,\bar\nu_2)}\\\hline
\mathrm{ph}&(\nu+\Omega,\nu';\nu,\nu'+\Omega)&\rev{(\nu_k+\Omega,\nu_k)}\\
\mathrm{pp}&(-\nu'-\Omega,\nu';\nu,-\nu-\Omega)&\rev{(\nu_k,-\nu_k-\Omega)}\\
\phbar&(\nu',\nu+\Omega;\nu,\nu'+\Omega)&\rev{(\nu_k,\nu_k+\Omega)}
\end{array}
\label{meth:leg-conventions}
\end{equation}
Writing $\Omega=\Omega_m$, $\nu=\nu_n$, and $\nu'=\nu_{n'}$\rev{, together
with the internal frequency $\nu_k$,} identifies the integer coordinates $(m,n,n',k)$ used
below. The identity $-\nu_n=\nu_{-n-1}$ supplies the offsets associated
with fermionic reversal.
\par

\par
For one orbital, the bare vertex amplitudes in the ordered spin basis
$(D,X)$ of Eq.~\eqref{main:spin-decomposition} are $(U,-U)$.
For two orbitals, define the
analogous orbital pairings \rev{$\theta_D,\theta_X$ and
$\theta_Y=\delta_{o_1o_2}\delta_{i_1i_2}$}, with these Kronecker deltas
acting on the orbital indices of the external legs.
The ordered basis and bare vertex are
\par
\begin{equation}
\begin{aligned}
 (T_1,\ldots,T_6)&=\rev{(\sigma_D\theta_D,\sigma_X\theta_X,\sigma_D\theta_X,}\\
 &\qquad\rev{\sigma_D\theta_Y,\sigma_X\theta_D,\sigma_X\theta_Y)},\\
 \boldsymbol\Gamma_0&=U(1,-1,0,0,0,0)\\
 &+J(-2,2,1,1,-1,-1).
\end{aligned}
\label{meth:flavor-basis}
\end{equation}
{The indices $a,b,c$ below enumerate basis components:
$D,X$ for one orbital and $1,\ldots,6$ for two degenerate orbitals.}
Contracting these tensors over the two internal flavors gives constant
coefficients {$C^{(r)}_{cab}$}. Their exact factorization
{$C^{(r)}_{cab}=\sum_\alpha O^{(r)}_{c\alpha}
L^{(r)}_{\alpha a}L^{(r)}_{\alpha b}$} reduces the frequency operation to
two terms per channel for one orbital and six for two orbitals. Identical
terms are combined before compression. {Supplementary Tables~S1 and S2 give the
exact flavor factors, and Supplementary Table~S3 lists the integer coordinate maps
(Supplementary Note~1).}

\subsection{Multiloop flow}
\label{meth:flow}

\par
We use the bubble contraction $\Bcal_r[A,P,B]$ defined in
Eq.~\eqref{main:bubble}, with flavor sums understood and
$(c_{\mathrm{ph}},c_{\mathrm{pp}},c_{\phbar})=(1,-1/2,-1)$;
the subscript includes the leg assignments in
Eq.~\eqref{meth:leg-conventions}.
For the recursive expressions below, we omit the scale index $\Lambda$.
We denote the first term in the self-energy flow of
Eq.~\eqref{main:integrated-flow} by
$\dot\Sigma_{\mathrm{std}}=\Gamma\mathbin{\cdot}S$ and use the Katanin
propagator derivative defined in Results.
The multiloop contributions with a bare fully irreducible vertex are
evaluated recursively\cite{KuglerGeneral2018}.
\par
Set {$\Pi_r=\rev{G(\bar\nu_1)G(\bar\nu_2)}$} and
$\dot I_r^{(\ell)}=\sum_{r'\ne r}\dot\gamma_{r'}^{(\ell)}$. Then
\begin{align}
 \dot\gamma_r^{(1)}&=\Bcal_r[\Gamma,\dot\Pi_r,\Gamma],\notag\\
 \dot\Pi_r&={\rev{\dot G_K(\bar\nu_1)G(\bar\nu_2)+G(\bar\nu_1)\dot G_K(\bar\nu_2)}},\notag\\
 \dot\gamma_{r,L}^{(\ell+1)}&=\Bcal_r[\dot I_r^{(\ell)},\Pi_r,\Gamma],\notag\\
 \dot\gamma_{r,R}^{(\ell+1)}&=\Bcal_r[\Gamma,\Pi_r,\dot I_r^{(\ell)}],\notag\\
 \dot\gamma_{r,C}^{(\ell+2)}&=\tfrac12\big[
 \Bcal_r[\Gamma,\Pi_r,\dot\gamma_{r,L}^{(\ell+1)}]+\Bcal_r[\dot\gamma_{r,R}^{(\ell+1)},\Pi_r,\Gamma]\big].
 \label{meth:loop-recursion}
\end{align}
\par
The last line averages the two equivalent central-term associations,
whose frequency restrictions differ on a finite grid. Its inserted block
belongs to channel $r$, not the complementary channels. At two loops only
the left and right terms occur. For $\ell\ge3$,
$\dot\gamma_r^{(\ell)}=\dot\gamma_{r,L}^{(\ell)}+
\dot\gamma_{r,R}^{(\ell)}+\dot\gamma_{r,C}^{(\ell)}$.
The complete derivative sums these orders through $L$.
\par

\par
The additional self-energy terms in Eq.~\eqref{main:integrated-flow} are
\begin{equation}
\begin{aligned}
 \dot\Sigma_{\mathrm c}&=
 \sum_{\ell=3}^L(\dot\gamma_{\mathrm{pp},C}^{(\ell)}+
 \dot\gamma_{\phbar,C}^{(\ell)})\mathbin{\cdot}G,\\
 \dot\Sigma_{\mathrm{fb}}&=\Gamma\mathbin{\cdot}
 (G\dot\Sigma_{\mathrm c}G).
\end{aligned}
\label{meth:self-energy-flow}
\end{equation}
\par
\par
The contraction of a vertex $A$ with a propagator $P$ is
\begin{equation}
 [A\mathbin{\cdot}P](\nu)=T\rev{\sum_{k,a}
 w_aA_a(\nu,\nu_k;\nu,\nu_k)P(\nu_k)},
 \label{meth:self-energy-contraction}
\end{equation}
with $w=(2,1)$ in the $(D,X)$ basis for one orbital and
$w=(4,1,2,2,2,1)$ in the ordered basis of
Eq.~\eqref{meth:flavor-basis} for two degenerate orbitals.
The $\mathrm{ph}$ central term is already generated by this
contraction. Within one derivative evaluation, $\dot G_K$ uses only
$\dot\Sigma_{\mathrm{std}}$; the additional self-energy terms are formed
after the vertex recursion, without a further iteration of the propagator
derivative.
\par

\subsection{Analytical tensor operators}
\label{meth:analytic-operators}

For $N=2^R$, $p=N/2$, and stored labels $n\in[-p,p-1]$, the offset code
is $x=n+p=\sum_{j=0}^{R-1}2^jb_j$. We group the bits of $D$ coordinates as
$s_j=1+\sum_{\mu=1}^D2^{\mu-1}b_{\mu j}$, ordered from least to most
significant. A local core {$T^{[j]}_{a,s,a'}$} has shape
$\chi_j\times2^D\times\chi_{j+1}$, with scalar open boundaries.
Self-energies, propagator pairs, vertices, and bubble intermediates have
$D=1,2,3,4$, respectively. The stored entry count is
$\sum_j2^D\chi_j\chi_{j+1}$\cite{Oseledets2011,Shinaoka2023}.

\par
A frequency transformation evaluates a source tensor at coordinates
$\boldsymbol y=\mathbf M\boldsymbol x+\boldsymbol b$ determined by the
target offset codes $\boldsymbol x$. For $D$ target and $D'$ source
coordinates, $\mathbf M\in\mathbb Z^{D'\times D}$. If the corresponding
map of physical integer labels has offset $\boldsymbol b_0$, then
$\boldsymbol b=\boldsymbol b_0+p\boldsymbol1_{D'}-
\mathbf M p\boldsymbol1_D$, where $\boldsymbol1_d$ is the length-$d$
vector of ones. Write
$b_\mu=\sum_{j\ge0}2^j\beta_{\mu j}$, with
$\beta_{\mu j}=\operatorname{sgn}(b_\mu)\operatorname{bit}_j(|b_\mu|)$.
\rev{Below, $x_{\alpha j},y_{\mu j}\in\{0,1\}$ denote the $j$th binary digits of
the offset codes $x_\alpha$ and $y_\mu$, and $\boldsymbol x_j,\boldsymbol y_j$
collect these digits at site $j$ [Fig.~\ref{fig:method}(b)]; they are not the
offset codes $\boldsymbol x,\boldsymbol y$ themselves.}
A carry
$\boldsymbol c_j$ specifies the local matrix product
operator\cite{Crosswhite2008,Kazeev2012}:
\par
\begin{equation}
\begin{aligned}
 z_{\mu j}&=c_{\mu j}+\beta_{\mu j}
 +\sum_\alpha M_{\mu\alpha}x_{\alpha j},\\
 W^{[j]}_{\boldsymbol c_j,\boldsymbol c_{j+1}}
 (\boldsymbol y_j,\boldsymbol x_j)&=
 \prod_\mu\delta_{y_{\mu j},z_{\mu j}\bmod2}
 \delta_{c_{\mu,j+1},(z_{\mu j}-y_{\mu j})/2},\\
 {\widetilde T^{[j]}_{(a,c),\boldsymbol x,(a',c')}}&=
 \sum_{\boldsymbol y}{T^{[j]}_{a,\boldsymbol y,a'}}
 W^{[j]}_{c,c'}(\boldsymbol y,\boldsymbol x).
\end{aligned}
\label{meth:carry-core}
\end{equation}
{The left boundary imposes $\boldsymbol c_0=0$.} \rev{The carries are integers,
not only $0$ or $1$. Because $x_{\alpha j}\in\{0,1\}$ and
$\beta_{\mu j}\in\{-1,0,1\}$, Eq.~\eqref{meth:carry-core} keeps
$c_{\mu j}$ within $[-1-\sum_\alpha\max(-M_{\mu\alpha},0),
\sum_\alpha\max(M_{\mu\alpha},0)]$, so each coordinate carries at most
$2+\sum_\alpha|M_{\mu\alpha}|$ values. The MPO bond dimension is therefore
bounded independently of $R$ by
$\prod_\mu(2+\sum_\alpha|M_{\mu\alpha}|)$; for the maps used here,
$|M_{\mu\alpha}|\le1$ (Supplementary Table~S3).} Remaining offset digits
are propagated at the right boundary with zero physical bits, accepting
only zero final carry. There is no periodic wrapping. Only allowed carry
transitions are stored. Reducible vertices vanish outside the source
domain; the bare interaction is added separately without this restriction.

The other analytical tensors are specified by their local cores in
Table~\ref{main:analytic-cores}. All omitted virtual dimensions and
boundaries are one. Concatenations act on virtual indices at fixed
physical index. Numerical MPS entries are obtained by factorization or
flow integration, not prescribed analytically.

\begin{table*}[t]
\centering\small
\caption{Analytical tensor constructions. Here $s=1+b_\Omega+2b_\nu+4b_{\nu'}+8b_k$ and $e=1+b_\Omega+2b_\nu+4b_{\nu'}$. {The bits $b_\Omega,b_\nu,b_{\nu'}$\rev{, and $b_k$} encode the integer labels $m,n,n'$\rev{, and $k$; the last labels the internal frequency $\nu_k$}.} \rev{Each entry states the input and output tensor trains.} Product virtual indices are written as pairs. {$A,B,Z$ are the factors in Eq.~\eqref{meth:pairwise}; self-energy closing uses vertex $A$ and propagator $P$, with external bit $b_e$.}}
\label{main:analytic-cores}
\begin{tabular}{p{.18\textwidth}p{.74\textwidth}}
\toprule
Operation & Local cores and boundaries\\\midrule
Constant $c$ & {$T^{[0]}_{1s1}=c$; $T^{[j]}_{1s1}=1$} for $j>0$.\\[3pt]
Exact sum {\rev{$\sum_\xi c_\xi\mathcal T_\xi$}} & \rev{Input: tensor trains $\mathcal T_\xi=(T_\xi^{[0]},\ldots,T_\xi^{[R-1]})$, $\xi=1,\ldots,K$, with common local index; output: their weighted sum $\mathcal T$.} {$T^{[0]}=(c_1T_1^{[0]}\ \cdots\ c_KT_K^{[0]})$};
 {$T^{[j]}=\operatorname{diag}(T_1^{[j]},\ldots,T_K^{[j]})$} for $0<j<R-1$;
 {$T^{[R-1]}=(T_1^{[R-1]};\ldots;T_K^{[R-1]})$}.
 For $R=1$, sum the single cores directly.\\[3pt]
Pair embedding & \rev{Input: two-frequency propagator pair $P$ with local index $t$ for $(b_\Omega,b_k)$; output: four-frequency tensor train $\widetilde P$ with local index $s$.} $E^{[j]}(s;t)=\delta_{t,1+b_\Omega+2b_k}$,
 $\widetilde P^{[j]}_{a,s,a'}=\sum_t E^{[j]}(s;t)P^{[j]}_{a,t,a'}$.
 One-frequency broadcasting uses $E(b_1,b_2;t)=\delta_{t,1+b_2}$.\\[3pt]
Internal sum & \rev{Input: four-frequency tensor train with local index $s$; output: three-frequency tensor train with local index $e$.} $S^{[j]}(e;s)=\delta_{e,1+b_\Omega+2b_\nu+4b_{\nu'}}$;
 both values of $b_k$ have weight one. Apply $c_rT$ once after the full sum.\\[3pt]
Elementwise product & $H^{[j]}_{(a,b),s,(a',b')}=
 \widetilde P^{[j]}_{a,s,a'}{B^{[j]}_{b,s,b'}}$.\\[3pt]
Summed product & $K^{[j]}_{(a,b),e,(a',b')}=
 \sum_{b_k}{A^{[j]}_{a,(e,b_k),a'}}Z^{[j]}_{b,(e,b_k),b'}$.\\[3pt]
Central restriction & From depth $R+1$ to $R$, leave sites $0,\ldots,R-2$
 unchanged and set {$\widetilde T^{[R-1]}_{a,\boldsymbol b,1}=
 \sum_tT^{[R-1]}_{a,\boldsymbol1-\boldsymbol b,t}
 T^{[R]}_{t,\boldsymbol b,1}$}.\\[3pt]
Self-energy closing & $C^{[j]}_{(a,b),b_e,(a',b')}=
 \sum_{b_k}{A^{[j]}_{a,(b_e,b_k),a'}P^{[j]}_{b,(b_e,b_k),b'}}$;
 external virtual boundaries are one and $T$ is applied after contraction.\\
\bottomrule
\end{tabular}
\end{table*}

\subsection{Propagators, pairwise contraction, and tensor rounding}
\label{meth:contractions}

The lower-temperature calculations assemble two-frequency propagator
pairs from one-frequency tensors. We sample $G$ and $\dot G_K$ at depth
$R+1$, because the shifted arguments $k+m$ and $-k-m-1$ exceed the stored
range. The analytic bare propagator is evaluated there with $\Sigma=0$
outside its stored domain. Affine operators construct the shifted factors;
the central restriction in Table~\ref{main:analytic-cores} returns them to
depth $R$ before their product is formed. This removes quadratic
two-frequency sampling, but still requires $O(N)$ one-frequency samples.
The earlier same-temperature reference and degenerate two-orbital computations
instead compressed densely sampled propagator pairs. Both use the same
subsequent compressed vertex contractions.

\par
Let $\mathcal R_{\epsilon,\chi}$ denote rounding with tolerance $\epsilon$
and maximum bond dimension $\chi$. After transforming and assembling the
source channels, $A_a(m,n,n',k)$ and $B_a(m,n,n',k)$ are the first and
second input amplitudes on a common four-frequency domain. Their source
arguments are given in Supplementary Table~S3. Form
$A_\alpha=\sum_a L_{\alpha a}A_a$ and
$B_\alpha=\sum_a L_{\alpha a}B_a$. The pairwise operation is
\par
\begin{equation}
\begin{aligned}
 Z_\alpha&=\mathcal R_{\epsilon/30,z}(\widetilde P\odot{B_\alpha}),\\
 Q_\alpha&=\mathcal R_{\epsilon,q}\left[\sum_k{A_\alpha}\odot Z_\alpha\right],\\
 {\bigl(\Bcal_r[A,P,B]\bigr)_c}&=\mathcal R_{\epsilon,\chi}
 \left[c_rT\sum_\alpha{O_{c\alpha}}Q_\alpha\right].
\end{aligned}
\label{meth:pairwise}
\end{equation}
\par
Here $\odot$ denotes elementwise multiplication. The limits $z$ and $q$
are the maximum bond dimensions before and after the internal-frequency
summation, respectively; $\chi$ limits the assembled output.
\par
\begin{revblock}
For each flavor factor $\alpha$, $Z_\alpha$ is formed in one left-to-right
sweep: the local cores of $\widetilde P$ and $B_\alpha$ are contracted site
by site, and the bond is truncated to at most $z$. $Z_\alpha$ is then
right-orthogonalized by a QR sweep without truncation, and $Q_\alpha$ is
formed by a second sweep that contracts $A_\alpha$ with $Z_\alpha$, sums the
internal bit $b_k$ at each site, and retains at most $q$. These steps
correspond to the inner loop of Algorithm~\ref{alg:flow}.
\end{revblock}
The local product and summed-product cores are in
Table~\ref{main:analytic-cores}. At each site, the accumulated left factor
is contracted into the local core, factorized, and passed onward. No
complete uncompressed four-frequency tensor train is formed. The two
products involve bond sizes {$r_Pr_B$} and {$r_Az$}, rather than retaining
{$r_Ar_Pr_B$} throughout{; $r_P,r_A,r_B$ denote the input bond dimensions}.

Canonical forms determine the meaning of the truncation weights. We
right-orthogonalize the two-coordinate propagator before embedding it.
The first left-to-right product is right-orthogonalized by a nontruncating
QR decomposition sweep before the second product. Source channels are first rounded
with $\epsilon/30$; their assembled factors use $\epsilon$. Flavor and
state sums are formed exactly before rounding. {For external
integer label $e$ and internal label $k$, self-energy closing uses source
triples $(0,e,k)$, $(-e-k-1,e,k)$, and $(k-e,e,e)$ for the $\mathrm{ph}$, $\mathrm{pp}$, and
$\phbar$ channels, respectively, followed by the closing core in
Table~\ref{main:analytic-cores}.}

\par
Ordinary rounding uses a left-to-right QR sweep followed by a
right-to-left singular-value decomposition (SVD). The smallest retained
rank satisfies $\|M-M_r\|_F^2\le\epsilon^2\mathcal N^2$ if permitted by
the prescribed maximum. Here $\|\cdot\|_F$ is the Frobenius norm and
$\mathcal N=\|M\|_F$ unless a positive external norm reference is supplied;
that reference remains fixed during the sweep. When the maximum is reached first, the
tolerance is not satisfied. The local criterion does not bound the
accumulated nonlinear-flow error. A full SVD is used when the smaller
matrix dimension is at most 32 or the tolerance is zero. Otherwise,
an adaptive Gaussian projection uses a dimension-dependent fixed random seed,
initial target dimension at most 64, and 16 additional vectors. The
target is doubled until the residual or dimension criterion is met.
For $M_{\mathrm{proj}}=Q^\dagger M$ and $M_{\mathrm{proj}}^\dagger=Q_1R_1$, an SVD of $R_1^\dagger$
avoids forming $M_{\mathrm{proj}}M_{\mathrm{proj}}^\dagger$. The residual includes the unprojected
weight. Right factors are normalized by their row norms; rows at the
numerical floor are set to zero. The resulting procedure is deterministic
for a fixed numerical environment, including at three and four loops.
In the $R=10$ altermagnetic flows, two arithmetic variants of these operations are used. Bubble terms are right-orthogonalized without truncation before they enter a sum, and a sum whose largest bond dimension exceeds $1.5(\chi+16)$ is first left-orthogonalized through a Gaussian tensor-train sketch of width $\chi+16$ with a shape-determined seed, after which the same truncating SVD sweep is applied. At $R=6$, these variants change the self-energies and static responses by at most $2.1\times10^{-4}$ relative to the procedure above, both in a complete one-loop flow and in a three-loop flow through 53 increments; in the complete one-loop flow this is 13 to 300 times smaller than the deviation of the truncated flow from an untruncated reference.
\par

\subsection{Scale integration and evaluation order}
\label{meth:integration}

We integrate in $s=\log(\Lambda_0/\Lambda)$ with
$f(s,y)=-\Lambda\dot y$ and
$y=(\Sigma,\gamma_{\mathrm{ph}},\gamma_{\mathrm{pp}},\gamma_{\phbar})$.
{Algorithm~\ref{alg:flow} specifies the derivative evaluation
before the integration procedure is described.}

{Transformed full vertices and propagator pairs without a scale derivative are reused}
within each derivative evaluation. Independent flavor contractions may
be evaluated concurrently, but their sum retains a fixed order.

\par
Initially $\Sigma=\gamma_r=0$, while $\Gamma_0$ is explicit. A Heun step
forms $f_0=f(s_0,y_0)$, $y_1^{\mathrm p}=y_0+hf_0$,
$f_1^{\mathrm p}=f(s_0+h,y_1^{\mathrm p})$, and
$y_1=y_0+h(f_0+f_1^{\mathrm p})/2$. Subsequently,
\begin{equation}
 \begin{aligned}
 y_{j+1}&=y_j+\tfrac h2(3\widehat f_j-\widehat f_{j-1}),\\
 \widehat f_0&=f_0,\qquad\widehat f_1=f_1^{\mathrm p},\\
 \widehat f_j&=f(s_j,y_j)\quad(j\ge2).
 \end{aligned}
\label{meth:ab2}
\end{equation}
Only the first Adams--Bashforth update reuses the predictor slope;
later slopes are evaluated at accepted states. Every state sum is rounded.
The initial predictor differs from the corrected state at order $h^2$,
consistent with second-order integration.
\par

\par
\begin{paperalgorithm}{Compressed flow derivative}
\label{alg:flow}
\begin{algorithmic}[1]
\Require State $y$, regulator $\Lambda$, maximum loop order $L$, tensor limits
\Ensure Logarithmic derivative $f(s,y)$
\Function{Contract}{$r,A,P,B$}
 \State \algwrap{Transform the selected source channels, reusing unchanged transforms}
 \State \algwrap{Round each source, assemble each factor, and round; add $\Gamma_0$ only to a full-vertex factor}
 \State \algwrap{Form the exact flavor combinations $A_\alpha,B_\alpha$}
 \State \algwrap{Right-orthogonalize $P$ and embed spectator bits}
 \For{each flavor factor $\alpha$}
 \State \algwrap{Contract $P$ with $B_\alpha$ site by site; retain $Z_\alpha$ with limit $z$}
 \State \algwrap{Right-orthogonalize $Z_\alpha$ without truncation}
 \State \algwrap{Contract $A_\alpha$ with $Z_\alpha$, summing the internal bit; retain limit $q$}
 \EndFor
 \State \algwrap{\Return the exact weighted flavor sum multiplied by $c_rT$, rounded to limit $\chi$}
\EndFunction
\State \algwrap{Evaluate $G,S$ and close $\Gamma$ with $S$ to obtain $\dot\Sigma_{\mathrm{std}}$}
\State \algwrap{Form $\dot G_K$ and the propagator pairs $\Pi_r,\dot\Pi_r$}
\State \algwrap{Call \Call{Contract}{$r,\Gamma,\dot\Pi_r,\Gamma$} for every one-loop channel}
\For{$\ell=2,\ldots,L$}
 \State \algwrap{Complete all preceding-order contributions before forming $\dot I_r^{(\ell-1)}$}
 \State \algwrap{Call \textsc{Contract} for the left and right terms in Eq.~\eqref{meth:loop-recursion}, with complementary-channel insertions}
 \If{$\ell\ge3$}
  \State \algwrap{Call \textsc{Contract} for both central associations, retaining the inserted channel $r$, and average}
 \EndIf
 \State \algwrap{Assemble each order in a fixed channel and flavor-summation order}
\EndFor
\State \algwrap{Sum channel derivatives through order $L$; form $\dot\Sigma_{\mathrm c},\dot\Sigma_{\mathrm{fb}}$}
\State \algwrap{\Return $-\Lambda(\dot\Sigma,\dot\gamma_{\mathrm{ph}},\dot\gamma_{\mathrm{pp}},\dot\gamma_{\phbar})$}
\end{algorithmic}
\end{paperalgorithm}
\par

\begin{table*}[t]
\centering\small
\caption{Numerical conditions for the reported calculations. All flows use logarithmic step $h=0.1$ and the pairwise contraction scheme.}
\label{main:parameter-table}
\begin{tabular}{lccccc}
\toprule
Calculation & $R$ & Loop orders & $T/U$ & $\epsilon$ & $\chi$\\\midrule
Single-orbital vertex and storage & $7,8,9,10,12$ & $1,2$ & $10^{-2}$ & $10^{-5}$ & $256$\\
Single-orbital responses & $12$ & $1,2$ & $10^{-2}$ & $10^{-5}$ & $256$\\
Single-orbital low-energy self-energy & $8$ & $1,2,3,4$ & $10^{-2}$ & $10^{-5}$ & $256$\\
Single-orbital low-temperature self-energy & $10$ & $1,2,3,4$ & $10^{-3}$ & $10^{-6}$ & $256$\\
Degenerate two-orbital response & $8$ & $1,2,3$ & $10^{-2}$ & $10^{-5}$ & $128$\\
{Altermagnetic self-energy and response} & {$10$} & {$1,2,3,4$} & $10^{-2}$ & $10^{-5}$ & $128$\\
{Altermagnetic supplementary comparisons} & {$7,8$} & {$1,2$} & $10^{-2}$ & $10^{-5}$ & $128$\\\bottomrule
\end{tabular}
\end{table*}
\subsection{Response functions and reference comparisons}
\label{main:response-method}

For the spin-symmetric model, set $\Pi_\nu(\Omega)=G(\ii\nu)G(\ii\nu+\ii\Omega)$
and $\Pi_0(\Omega)=T\sum_\nu\Pi_\nu(\Omega)$; the subscript zero here
denotes the frequency sum. The connected density response is
\begin{equation}
 \chi_{\sigma\sigma'}(\ii\Omega)=
 -\delta_{\sigma\sigma'}\Pi_0
 -T^2\sum_{\nu,\nu'}\Pi_\nu
 \Gamma_{\sigma\sigma'}(\Omega;\nu,\nu')\Pi_{\nu'}.
\label{main:response-full}
\end{equation}
{Here $\Gamma_{\sigma\sigma'}$ denotes the full vertex with
outgoing and incoming spins $(\sigma,\sigma';\sigma,\sigma')$ at the
$\mathrm{ph}$ frequencies of Eq.~\eqref{meth:leg-conventions}.}
All channels and the bare interaction enter the full vertex in the same
{$\mathrm{ph}$ coordinates}. A further Bethe--Salpeter iteration would
count these contributions again. The NRG normalization is
$\overline\chi_a=(\Delta_{\mathrm{eff}}/4)
\operatorname{Re}\sum_{\sigma\sigma'}w^a_\sigma w^a_{\sigma'}\chi_{\sigma\sigma'}$,
with $w^{\mathrm d}=(1,1)$ and $w^{\mathrm m}=(1,-1)$.

{For two degenerate orbitals, the weights that combine
the full vertex amplitudes into charge, spin, and orbital-quadrupole
components in the basis of Eq.~\eqref{meth:flavor-basis} are}
$\mathbf q_{\mathrm{ch}}=(4,1,2,2,2,1)$,
$\mathbf q_{\mathrm{sp}}=(0,1,0,0,2,1)$, and
$\mathbf q_{\mathrm{orb}}=(0,1,2,2,0,1)$.
Writing $F_k=\mathbf q_k\cdot\boldsymbol\Gamma$ gives
$\chi_k=-4c_k[\Pi_0+T^2\sum_{\nu\nu'}\Pi_\nu F_k\Pi_{\nu'}]$,
where $c_{\mathrm{ch}}=1$ and $c_{\mathrm{sp}}=c_{\mathrm{orb}}=1/4$.
The denominator in Fig.~\ref{fig:hund}(b) is the dressed bubble
$-4c_k\Pi_0$. {The interaction components in
Fig.~\ref{fig:hund}(a) are
$g_k=\operatorname{Re}F_k(0;\pi T,-\pi T)$,
using the full vertex in the same $\mathrm{ph}$ coordinates.}

The PA and NRG comparisons use the Matsubara reference
data\cite{Ge2024,Ge2023}; we perform no new NRG calculation.
The PA channel labels $(t,p,a)$ map to $(\mathrm{ph},\mathrm{pp},\phbar)$,
and its overall vertex sign is opposite. The complete $256\times256$
planes compare {$Q_r=\operatorname{Re}\gamma_{r,D}/U$} with
$P_r=-\operatorname{Re}\phi_r^{\mathrm{PA}}$, {where
$\phi_r^{\mathrm{PA}}$ is the channel-reducible vertex of the reference,
already normalized by $U$}.
Their combined deviation is
$[\sum_r\|Q_r-P_r\|_F^2/\sum_r\|P_r\|_F^2]^{1/2}$,
without individual channel rescaling. NRG susceptibilities already carry
the reference normalization and require only sign reversal. Relative
response and low-energy self-energy errors use the root-mean-square
pointwise relative deviation over the lowest eight frequencies.
Matsubara indices are matched, retaining the small displacement between
$u=0.99970$ and the reference $u=1$ in the frequency axes. The NRG Hartree
term does not affect the imaginary self-energy.

\subsection{Self-energy subtraction and analytic continuation}
\label{main:self-energy-method}

\par
Figure~\ref{fig:siam-dynamics}(c) plots the residual
$\mathcal R_n$ of Eq.~\eqref{main:sigma-residual} for $n=1,\ldots,7$,
normalized by $\Delta_{\mathrm{eff}}$, against
$[\nu_n^2-(\pi T)^2]/\Delta_{\mathrm{eff}}^2$.
\par

\rev{Analytic continuation is used only for Fig.~\ref{fig:siam-dynamics}(d); the
real-frequency hybridization widths in Fig.~\ref{fig:am-combined}(b) follow
directly from Eq.~\eqref{am:compact-bath}.}
The low-temperature real-frequency {calculations} use the standard
continued-fraction Pad\'e construction\cite{Vidberg1977} from the first
80 positive Matsubara self-energy values, starting at $\pi T$.
The rational function is evaluated at
{$(\omega+\ii\eta_{\mathrm{cont}})/\Delta_{\mathrm{eff}}$,
with positive continuation broadening
$\eta_{\mathrm{cont}}/\Delta_{\mathrm{eff}}=10^{-4}$}; numerator and denominator are
rescaled together to avoid overflow. Set
$s_\ell(x)=-\operatorname{Im}\Sigma^{\mathrm{ret}}_\ell(x\Delta_{\mathrm{eff}})
/\Delta_{\mathrm{eff}}$ and subtract its mean over
$x=0.0035,0.005,0.007$, denoted $\overline c_\ell$.
\par
Figure~\ref{fig:siam-dynamics}(d) displays $y_4=s_4-\overline c_4$;
its dimensional constant is $c_T=\Delta_{\mathrm{eff}}\overline c_4$.
The quadratic guide is $y_4(0.05)(x/0.05)^2$, using linear interpolation
at $x=0.05$. Only its position, not its exponent, follows the data.
\par
Using 40--100 input points preserves the approximately quadratic
dependence on $0.01\le x\le0.2$. This does not determine the full
continuation uncertainty or identify the subtracted constant as a
thermal scattering rate.

\subsection{Altermagnetic impurity and observables}
\label{am:methods-compact}

\par
Integrating out the host of Eq.~\eqref{am:host-hamiltonian} gives the
hybridization matrix at complex frequency $z$,
\begin{equation}
 \Delta_{mm',\sigma}(z)=\int_{[-\pi,\pi]^2}\frac{d^2k}{(2\pi)^2}
 \frac{V_{m,\mathbf k}V_{m',\mathbf k}^*}{z-\epsilon_{\mathbf k\sigma}}.
 \label{am:compact-bath}
\end{equation}
\par
Reflection symmetry removes off-diagonal elements. With flavor order $(xz\uparrow,yz\uparrow,xz\downarrow,yz\downarrow)$, $s_f=(+,-,-,+)$ labels the two diagonal functions. We fix $t=4\Delta_{\mathrm{ref}}$, $\Delta_{\mathrm{ref}}=4V^2/(\pi t)$, and retain $V$ when varying $t_J$. The calculations use $t_J/t=0.3$, $U=1.5\pi\Delta_{\mathrm{ref}}$, and $T/U=10^{-2}$. {The self-energy and orbital-field responses in Fig.~\ref{fig:am-combined}(c,d) use $R=10$, one to four loops, and $J/U=0,0.10,0.15,0.20,0.25,0.30$ with the tensor parameters given below; Supplementary Note~8 quantifies their loop-order dependence. The regulator comparison in Supplementary Fig.~S4 uses $R=7$, and additional one-loop comparisons at $R=7,8$ are presented in Supplementary Note~5.} Momentum-grid refinement resolves $\Delta_s(i\nu)$ to relative tolerance $10^{-8}$ through $n_{\max}=2^{R+1}$. With $\Gamma_s(\omega)=-\operatorname{Im}\Delta_s(\omega+i0^+)$, \rev{the extension to $|\nu|>\nu_{n_{\max}}$} is
\begin{align}
 \Delta_s(i\nu)&=-i\operatorname{sgn}(\nu)\widetilde\Gamma_s(|\nu|),\nonumber\\
 \widetilde\Gamma_s(\nu)&=\frac{M_{0s}}\nu-\frac{M_{2s}}{\nu^3}
 +\frac{M_{4s}}{\nu^5}+\frac{a_{7s}}{\nu^7},\nonumber\\
 M_{2q,s}&=\int\frac{d\omega}{\pi}\,\Gamma_s(\omega)\omega^{2q}.
 \label{am:compact-bath-tail}
\end{align}
The coefficient $a_{7s}$ imposes continuity at $\nu_{n_{\max}}$, above the band edge.

The 24 flavor tensors have a compact constructive definition. Starting with the six tensors of Eq.~\eqref{meth:flavor-basis}, visit {$b=1,\ldots,6$} and the four-sign tuples $p$ in lexicographic order, with $+$ before $-$. Append {$T_b\Pi^p$} only when linearly independent of the retained tensors\rev{; this retains 4, 4, 8, 4, 2, and 2 tensors for $b=1,\ldots,6$, including the unprojected ones; the resulting 24-dimensional space contains the bare vertex and is closed under the flavor contractions below,} where
\begin{equation}
 \Pi^p_{12;34}=\prod_{j=1}^4\frac{1+p_js_{f_j}}2.
 \label{am:compact-projector}
\end{equation}
{For each ordered internal sign pair $\kappa=(s_i,s_j)$, with $s_i,s_j=\pm1$, the flavor contractions are}
\begin{align}
 \mathcal C_{\mathrm{ph}}^\kappa({A,B})_{12;34}
 &=\sum_{i,j:(s_i,s_j)=\kappa}{A_{i2;j4}B_{1j;3i}},\nonumber\\
 \mathcal C_{\mathrm{pp}}^\kappa({A,B})_{12;34}
 &=\sum_{i,j:(s_i,s_j)=\kappa}{A_{12;ij}B_{ij;34}},\nonumber\\
 \mathcal C_{\phbar}^\kappa({A,B})_{12;34}
 &=\sum_{i,j:(s_i,s_j)=\kappa}{A_{1j;i4}B_{i2;3j}}.
 \label{am:compact-flavor-contractions}
\end{align}
{For this basis, the component indices $a,b,c$ run from 1 to 24.}
Their basis coefficients factor exactly into 12 terms {$\sum_\alpha O_{c\alpha}L_{\alpha a}R_{\alpha b}$, with output component $c$ and input components $a,b$}; left and right factors generally differ. Channel signs follow the common channel factors above. Self-energy weights are {$w_a(s,s')=\sum_{g:s_g=s'}T_{a,fg;fg}$} for $s_f=s$.

\par
We use $G_s^{-1}=i\nu-\Delta_s+i\Lambda\operatorname{sgn}\nu-\Sigma_s$ and the common Heun--Adams--Bashforth integration with step $0.1$, evaluating the displayed observables after 76 increments from $\Lambda_0=1000\Delta_{\mathrm{ref}}$. One loop includes Katanin feedback. Higher loops follow Eqs.~\eqref{meth:loop-recursion} and \eqref{meth:self-energy-flow}, with the sign-resolved propagators, distinct first- and second-input flavor factors, and self-energy weights defined above. The three- and four-loop calculations include the central vertex contributions and their self-energy feedback.

The tolerance is $\epsilon=10^{-5}$, with maximum bond dimensions $\chi=128$ for the state and $z=q=256$ for the intermediate products. All field norms below are Frobenius norms over the represented frequency grid. Vertex components in a state update share the largest vertex-component norm of the accepted state as their rounding reference. Higher-loop derivative sums instead use the largest vertex-component norm of the one-loop derivative. Before transformation, an input amplitude is represented by zero if it has no included nonzero bare contribution and every included source-channel component has norm at most $10^{-14}$ times the largest vertex-component norm of that source state.

Cancellation among flavor amplitudes determines the rounding reference for some bubble terms. For a rounded factor $F$ assembled from transformed amplitudes $F_a$ and coefficients $v_a$, the criterion is
\begin{equation}
 \|F\|_F\le 10^3\epsilon\sum_a |v_a|\,\|F_a\|_F,
 \label{am:norm-reference}
\end{equation}
applied only when more than one coefficient is nonzero. Within each sign class, bubbles for which neither factor satisfies this criterion are contracted first. Their largest norm provides the reference for the remaining bubbles: the second, internally summed product is rounded using $10^{-2}$ times that norm. If it vanishes, the local matrix norm is used. This criterion changes the rounding reference but does not omit a nonzero contraction.

After all sign classes are evaluated, each nonzero bubble term smaller than $10^{-2}\epsilon$ times the largest term norm in the channel is rounded again against that largest norm, with limit $\chi$. Weighted flavor sums are rounded within each sign class before the classes are combined and rounded. An assembled channel component is represented by zero when its norm is at most $10^{-14}$ times the largest weighted sum of constituent term norms among that channel's output components. This reference uses absolute contraction weights, including the prefactor $c_rT$.
\par

\par
{Let $d$ be the column of annihilation operators in the stated spin and orbital basis. A one-body observable $\hat O=d^\dagger\mathbf O d$ is specified by a $4\times4$ matrix $\mathbf O$ with elements $(\mathbf O)_{fg}$. We use $\mathbf O_n=I_4$, $\mathbf O_{S^z}=\boldsymbol\sigma^z\otimes I_2/2$, $\mathbf O_{\tau^z}=I_2\otimes\boldsymbol\tau^z/2$, and $\mathbf O_{O_\alpha}=\boldsymbol\sigma^\alpha\otimes\boldsymbol\tau^z/4$ ($\alpha=x,z$). Here $\boldsymbol\sigma^\alpha$ and $\boldsymbol\tau^\alpha$ are spin and orbital Pauli matrices, respectively, and $I_j$ is the $j\times j$ identity matrix. The last matrices define $\hat O_\alpha=d^\dagger\mathbf O_{O_\alpha}d$.}
For $G_f(n)=G_{s_f}(i\nu_n)$, set $P_{fg}(n)=G_f(n+m)G_g(n)$ and $\bar P_{hk}(n')=G_h(n')G_k(n'+m)$. {The connected autocorrelation of $\hat O$ is}
\par
\begin{align}
 {\chi_{O,O}}(i\Omega_m)&=-T\sum_{nfg}{(\mathbf O)_{fg}(\mathbf O)_{gf}}P_{fg}(n)\nonumber\\
 &\quad-T^2\sum_{nn'fghk}{(\mathbf O)_{fg}(\mathbf O)_{hk}}P_{fg}(n)\nonumber\\
 &\quad\times\Gamma_{fh;gk}(n+m,n';n,n'+m)\bar P_{hk}(n').
 \label{am:compact-response}
\end{align}
The full vertex includes the bare interaction and three channels; both fermionic sums use $[-2^{R-1},2^{R-1}-1]$. No additional Bethe--Salpeter iteration is performed. The normalization ${\chi_{O,O}^{(0)}}$ uses $U=J=0$ at the same bath, temperature, regulator, and frequency range. We define $\delta\Sigma=\operatorname{Im}(\Sigma_+-\Sigma_-)$, $\overline{\delta\Sigma}=\frac17\sum_{n=2}^8\delta\Sigma(i\nu_n)$\rev{, which excludes the lowest Matsubara frequencies; there the difference of two nearly equal self-energies is most sensitive to truncation errors, as measured by violations of the exact odd-frequency relation $\delta\Sigma(i\nu_n)+\delta\Sigma(i\nu_{-n-1})=0$}, and {$\eta=\chi_{O_z,O_z}(0)-\chi_{O_x,O_x}(0)$. These last two susceptibilities are autocorrelations of $\hat O_z$ and $\hat O_x$, not the mixed response $\chi_{S^z,\tau^z}$.}

{For a second observable $\hat P=d^\dagger\mathbf P d$, the mixed susceptibility $\chi_{O,P}$ follows from Eq.~\eqref{am:compact-response} by replacing $(\mathbf O)_{gf}$ and $(\mathbf O)_{hk}$ with $(\mathbf P)_{gf}$ and $(\mathbf P)_{hk}$ in the first and second terms, respectively. The bold matrix $\mathbf P$ differs from the propagator product $P_{fg}(n)$. The orbital-field response uses $\mathbf O=\mathbf O_{S^z}$ and $\mathbf P=\mathbf O_{\tau^z}$ in the numerator of Eq.~\eqref{am:orbital-polarization}, and $\mathbf O=\mathbf P=\mathbf O_{\tau^z}$ in its denominator.} Both are evaluated at zero transfer from the same unperturbed state. Its dressed-bubble counterpart retains only the first term of each contraction, using the dressed propagators of the corresponding loop order at the same regulator and frequency range. It is distinct from the noninteracting normalization used for the supplementary spin and orbital susceptibilities. The supplementary regulator comparison continues the two-loop flows at $J/U=0.10,0.20,0.30$ to 92 and 100 increments, reaching $\Lambda/\Delta_{\mathrm{ref}}\simeq0.101$ and $0.0454$, respectively; all other parameters and response definitions are unchanged.

In the $R=10$ flows the bond dimension of the vertex state remains below the prescribed $\chi=128$, reaching at most 119 during the flow and 118 at the endpoint, whereas the intermediate limits $z=q=256$ are reached in every flow. At $R=7$ and $J/U=0.30$, doubling $\chi$, $z$, and $q$ changes the one-loop $\overline{\delta\Sigma}$ by $0.14\%$; this one-loop comparison does not establish bond-dimension convergence at higher loop orders or at $R=10$. Figure~\ref{fig:am-combined}(d) establishes the persistence of the Hund dependence at the stated numerical parameters; convergence in loop order has not been established, since $\overline{\delta\Sigma}$ changes by up to $14\%$ and the polarization at $J/U=0.30$ by $18\%$ between three and four loops, whereas the polarization at $J/U\le0.15$ changes by less than $4\%$ (Supplementary Note~8). In the one-loop calculations at $R=7$ and $8$, the anisotropy $\eta$ is more sensitive than its individual susceptibilities, and nonmonotonic charge-response data are excluded from physical conclusions (Supplementary Note~5).

\subsection{Numerical conditions and validation}
\label{main:validation}

For the flat-bath models we set $\Delta=1$, $U=3\pi/2$,
$\Lambda_0=1000$, and $h=0.1$; 76 increments reach
$\Lambda=0.500$. The tensor parameters are summarized in
Table~\ref{main:parameter-table}. These are prescribed maxima, not the
dimensions attained at every bond. The intermediate limits are
$z=q=2\chi$ and their first-product tolerance is $\epsilon/30$.

\par
Integer maps, flavor contractions, uncompressed frequency sums, and
noninteracting and bare-vertex identities (Supplementary Note~2)
independently reproduce the prescribed algebra to machine precision.
The one-loop endpoint opposite-spin vertex changes by less than $0.05\%$
between $\chi=192$ and $256$ over $R=6,\ldots,10$; this does not establish
higher-loop or pointwise vertex convergence. Reducing the integration
step adds rounding operations, so its effect need not isolate integration error.

Calculations use Julia and complex double precision. The reference result
uses six Julia threads and one linear-algebra thread, with fixed summation
order and projection seeds. Selected high-rank derivatives and restarted
segments, including the two-loop $R=7,\ldots,10$ calculations, were reproduced
in independent processes. Storage for the QTT representation is calculated from actual complex entries;
dense estimates use six vertex amplitudes and 16 bytes per entry. Neither
includes temporary contraction arrays or measures runtime. A common
global tolerance does not imply equal low-frequency errors at different $R$.

An AI assistant (Claude, Anthropic) was used to help correct errors in the simulation code and to assist with data analysis. The authors verified all code changes and numerical results and take full responsibility for the content.
\par

\section*{Data and code availability}
The independent NRG and parquet reference data are available in the repository associated with the reference study\cite{Ge2023}. Source data underlying the figures are available from the corresponding authors upon reasonable request. The source code for the single-orbital and degenerate two-orbital calculations is available at \url{https://github.com/hylee-ku/qtt-frg}. The code for the altermagnetic calculations is available from the corresponding authors upon reasonable request.

\begin{acknowledgments}
H.-Y.L. was supported by the Basic Science Research Program through the National Research Foundation of Korea funded by the Ministry of Science and ICT [Grant Nos. RS-2023-00220471 and RS-2025-16064392]. He also acknowledges the hospitality of the Institute for Solid State Physics, The University of Tokyo, where part of this work was carried out. N.K. acknowledges support from JSPS KAKENHI Grant No.~23K25789.
\end{acknowledgments}

\bibliography{refs}

\begin{thebibliography}{38}%
\makeatletter
\providecommand \@ifxundefined [1]{%
 \@ifx{#1\undefined}
}%
\providecommand \@ifnum [1]{%
 \ifnum #1\expandafter \@firstoftwo
 \else \expandafter \@secondoftwo
 \fi
}%
\providecommand \@ifx [1]{%
 \ifx #1\expandafter \@firstoftwo
 \else \expandafter \@secondoftwo
 \fi
}%
\providecommand \natexlab [1]{#1}%
\providecommand \enquote  [1]{``#1''}%
\providecommand \bibnamefont  [1]{#1}%
\providecommand \bibfnamefont [1]{#1}%
\providecommand \citenamefont [1]{#1}%
\providecommand \href@noop [0]{\@secondoftwo}%
\providecommand \href [0]{\begingroup \@sanitize@url \@href}%
\providecommand \@href[1]{\@@startlink{#1}\@@href}%
\providecommand \@@href[1]{\endgroup#1\@@endlink}%
\providecommand \@sanitize@url [0]{\catcode `\\12\catcode `\$12\catcode
  `\&12\catcode `\#12\catcode `\^12\catcode `\_12\catcode `\%12\relax}%
\providecommand \@@startlink[1]{}%
\providecommand \@@endlink[0]{}%
\providecommand \url  [0]{\begingroup\@sanitize@url \@url }%
\providecommand \@url [1]{\endgroup\@href {#1}{\urlprefix }}%
\providecommand \urlprefix  [0]{URL }%
\providecommand \Eprint [0]{\href }%
\providecommand \doibase [0]{https://doi.org/}%
\providecommand \selectlanguage [0]{\@gobble}%
\providecommand \bibinfo  [0]{\@secondoftwo}%
\providecommand \bibfield  [0]{\@secondoftwo}%
\providecommand \translation [1]{[#1]}%
\providecommand \BibitemOpen [0]{}%
\providecommand \bibitemStop [0]{}%
\providecommand \bibitemNoStop [0]{.\EOS\space}%
\providecommand \EOS [0]{\spacefactor3000\relax}%
\providecommand \BibitemShut  [1]{\csname bibitem#1\endcsname}%
\let\auto@bib@innerbib\@empty
\bibitem [{\citenamefont {Georges}\ \emph {et~al.}(1996)\citenamefont
  {Georges}, \citenamefont {Kotliar}, \citenamefont {Krauth},\ and\
  \citenamefont {Rozenberg}}]{Georges1996}%
  \BibitemOpen
  \bibfield  {author} {\bibinfo {author} {\bibfnamefont {A.}~\bibnamefont
  {Georges}}, \bibinfo {author} {\bibfnamefont {G.}~\bibnamefont {Kotliar}},
  \bibinfo {author} {\bibfnamefont {W.}~\bibnamefont {Krauth}},\ and\ \bibinfo
  {author} {\bibfnamefont {M.~J.}\ \bibnamefont {Rozenberg}},\ }\href
  {https://doi.org/10.1103/RevModPhys.68.13} {\bibfield  {journal} {\bibinfo
  {journal} {Rev. Mod. Phys.}\ }\textbf {\bibinfo {volume} {68}},\ \bibinfo
  {pages} {13} (\bibinfo {year} {1996})}\BibitemShut {NoStop}%
\bibitem [{\citenamefont {Metzner}\ \emph {et~al.}(2012)\citenamefont
  {Metzner}, \citenamefont {Salmhofer}, \citenamefont {Honerkamp},
  \citenamefont {Meden},\ and\ \citenamefont {Sch{\"o}nhammer}}]{Metzner2012}%
  \BibitemOpen
  \bibfield  {author} {\bibinfo {author} {\bibfnamefont {W.}~\bibnamefont
  {Metzner}}, \bibinfo {author} {\bibfnamefont {M.}~\bibnamefont {Salmhofer}},
  \bibinfo {author} {\bibfnamefont {C.}~\bibnamefont {Honerkamp}}, \bibinfo
  {author} {\bibfnamefont {V.}~\bibnamefont {Meden}},\ and\ \bibinfo {author}
  {\bibfnamefont {K.}~\bibnamefont {Sch{\"o}nhammer}},\ }\href
  {https://doi.org/10.1103/RevModPhys.84.299} {\bibfield  {journal} {\bibinfo
  {journal} {Rev. Mod. Phys.}\ }\textbf {\bibinfo {volume} {84}},\ \bibinfo
  {pages} {299} (\bibinfo {year} {2012})}\BibitemShut {NoStop}%
\bibitem [{\citenamefont {Georges}\ \emph {et~al.}(2013)\citenamefont
  {Georges}, \citenamefont {de' Medici},\ and\ \citenamefont
  {Mravlje}}]{Georges2013}%
  \BibitemOpen
  \bibfield  {author} {\bibinfo {author} {\bibfnamefont {A.}~\bibnamefont
  {Georges}}, \bibinfo {author} {\bibfnamefont {L.}~\bibnamefont {de'
  Medici}},\ and\ \bibinfo {author} {\bibfnamefont {J.}~\bibnamefont
  {Mravlje}},\ }\href
  {https://doi.org/10.1146/annurev-conmatphys-020911-125045} {\bibfield
  {journal} {\bibinfo  {journal} {Annu. Rev. Condens. Matter Phys.}\ }\textbf
  {\bibinfo {volume} {4}},\ \bibinfo {pages} {137} (\bibinfo {year}
  {2013})}\BibitemShut {NoStop}%
\bibitem [{\citenamefont {Anderson}(1961)}]{Anderson1961}%
  \BibitemOpen
  \bibfield  {author} {\bibinfo {author} {\bibfnamefont {P.~W.}\ \bibnamefont
  {Anderson}},\ }\href {https://doi.org/10.1103/PhysRev.124.41} {\bibfield
  {journal} {\bibinfo  {journal} {Phys. Rev.}\ }\textbf {\bibinfo {volume}
  {124}},\ \bibinfo {pages} {41} (\bibinfo {year} {1961})}\BibitemShut
  {NoStop}%
\bibitem [{\citenamefont {Bulla}\ \emph {et~al.}(2008)\citenamefont {Bulla},
  \citenamefont {Costi},\ and\ \citenamefont {Pruschke}}]{Bulla2008}%
  \BibitemOpen
  \bibfield  {author} {\bibinfo {author} {\bibfnamefont {R.}~\bibnamefont
  {Bulla}}, \bibinfo {author} {\bibfnamefont {T.~A.}\ \bibnamefont {Costi}},\
  and\ \bibinfo {author} {\bibfnamefont {T.}~\bibnamefont {Pruschke}},\ }\href
  {https://doi.org/10.1103/RevModPhys.80.395} {\bibfield  {journal} {\bibinfo
  {journal} {Rev. Mod. Phys.}\ }\textbf {\bibinfo {volume} {80}},\ \bibinfo
  {pages} {395} (\bibinfo {year} {2008})}\BibitemShut {NoStop}%
\bibitem [{\citenamefont {Wentzell}\ \emph {et~al.}(2020)\citenamefont
  {Wentzell}, \citenamefont {Li}, \citenamefont {Tagliavini}, \citenamefont
  {Taranto}, \citenamefont {Rohringer}, \citenamefont {Held}, \citenamefont
  {Toschi},\ and\ \citenamefont {Andergassen}}]{Wentzell2020}%
  \BibitemOpen
  \bibfield  {author} {\bibinfo {author} {\bibfnamefont {N.}~\bibnamefont
  {Wentzell}}, \bibinfo {author} {\bibfnamefont {G.}~\bibnamefont {Li}},
  \bibinfo {author} {\bibfnamefont {A.}~\bibnamefont {Tagliavini}}, \bibinfo
  {author} {\bibfnamefont {C.}~\bibnamefont {Taranto}}, \bibinfo {author}
  {\bibfnamefont {G.}~\bibnamefont {Rohringer}}, \bibinfo {author}
  {\bibfnamefont {K.}~\bibnamefont {Held}}, \bibinfo {author} {\bibfnamefont
  {A.}~\bibnamefont {Toschi}},\ and\ \bibinfo {author} {\bibfnamefont
  {S.}~\bibnamefont {Andergassen}},\ }\href
  {https://doi.org/10.1103/PhysRevB.102.085106} {\bibfield  {journal} {\bibinfo
   {journal} {Phys. Rev. B}\ }\textbf {\bibinfo {volume} {102}},\ \bibinfo
  {pages} {085106} (\bibinfo {year} {2020})}\BibitemShut {NoStop}%
\bibitem [{\citenamefont {Bonetti}\ \emph {et~al.}(2022)\citenamefont
  {Bonetti}, \citenamefont {Toschi}, \citenamefont {Hille}, \citenamefont
  {Andergassen},\ and\ \citenamefont {Vilardi}}]{Bonetti2022}%
  \BibitemOpen
  \bibfield  {author} {\bibinfo {author} {\bibfnamefont {P.~M.}\ \bibnamefont
  {Bonetti}}, \bibinfo {author} {\bibfnamefont {A.}~\bibnamefont {Toschi}},
  \bibinfo {author} {\bibfnamefont {C.}~\bibnamefont {Hille}}, \bibinfo
  {author} {\bibfnamefont {S.}~\bibnamefont {Andergassen}},\ and\ \bibinfo
  {author} {\bibfnamefont {D.}~\bibnamefont {Vilardi}},\ }\href
  {https://doi.org/10.1103/PhysRevResearch.4.013034} {\bibfield  {journal}
  {\bibinfo  {journal} {Phys. Rev. Research}\ }\textbf {\bibinfo {volume}
  {4}},\ \bibinfo {pages} {013034} (\bibinfo {year} {2022})}\BibitemShut
  {NoStop}%
\bibitem [{\citenamefont {Shinaoka}\ \emph {et~al.}(2023)\citenamefont
  {Shinaoka}, \citenamefont {Wallerberger}, \citenamefont {Murakami},
  \citenamefont {Nogaki}, \citenamefont {Sakurai}, \citenamefont {Werner},\
  and\ \citenamefont {Kauch}}]{Shinaoka2023}%
  \BibitemOpen
  \bibfield  {author} {\bibinfo {author} {\bibfnamefont {H.}~\bibnamefont
  {Shinaoka}}, \bibinfo {author} {\bibfnamefont {M.}~\bibnamefont
  {Wallerberger}}, \bibinfo {author} {\bibfnamefont {Y.}~\bibnamefont
  {Murakami}}, \bibinfo {author} {\bibfnamefont {K.}~\bibnamefont {Nogaki}},
  \bibinfo {author} {\bibfnamefont {R.}~\bibnamefont {Sakurai}}, \bibinfo
  {author} {\bibfnamefont {P.}~\bibnamefont {Werner}},\ and\ \bibinfo {author}
  {\bibfnamefont {A.}~\bibnamefont {Kauch}},\ }\href
  {https://doi.org/10.1103/PhysRevX.13.021015} {\bibfield  {journal} {\bibinfo
  {journal} {Phys. Rev. X}\ }\textbf {\bibinfo {volume} {13}},\ \bibinfo
  {pages} {021015} (\bibinfo {year} {2023})}\BibitemShut {NoStop}%
\bibitem [{\citenamefont {Rohshap}\ \emph {et~al.}(2025)\citenamefont
  {Rohshap}, \citenamefont {Ritter}, \citenamefont {Shinaoka}, \citenamefont
  {von Delft}, \citenamefont {Wallerberger},\ and\ \citenamefont
  {Kauch}}]{Rohshap2024}%
  \BibitemOpen
  \bibfield  {author} {\bibinfo {author} {\bibfnamefont {S.}~\bibnamefont
  {Rohshap}}, \bibinfo {author} {\bibfnamefont {M.~K.}\ \bibnamefont {Ritter}},
  \bibinfo {author} {\bibfnamefont {H.}~\bibnamefont {Shinaoka}}, \bibinfo
  {author} {\bibfnamefont {J.}~\bibnamefont {von Delft}}, \bibinfo {author}
  {\bibfnamefont {M.}~\bibnamefont {Wallerberger}},\ and\ \bibinfo {author}
  {\bibfnamefont {A.}~\bibnamefont {Kauch}},\ }\href
  {https://doi.org/10.1103/PhysRevResearch.7.023087} {\bibfield  {journal}
  {\bibinfo  {journal} {Phys. Rev. Research}\ }\textbf {\bibinfo {volume}
  {7}},\ \bibinfo {pages} {023087} (\bibinfo {year} {2025})}\BibitemShut
  {NoStop}%
\bibitem [{\citenamefont {Wetterich}(1993)}]{Wetterich1993}%
  \BibitemOpen
  \bibfield  {author} {\bibinfo {author} {\bibfnamefont {C.}~\bibnamefont
  {Wetterich}},\ }\href {https://doi.org/10.1016/0370-2693(93)90726-X}
  {\bibfield  {journal} {\bibinfo  {journal} {Phys. Lett. B}\ }\textbf
  {\bibinfo {volume} {301}},\ \bibinfo {pages} {90} (\bibinfo {year}
  {1993})}\BibitemShut {NoStop}%
\bibitem [{\citenamefont {Salmhofer}\ and\ \citenamefont
  {Honerkamp}(2001)}]{Salmhofer2001}%
  \BibitemOpen
  \bibfield  {author} {\bibinfo {author} {\bibfnamefont {M.}~\bibnamefont
  {Salmhofer}}\ and\ \bibinfo {author} {\bibfnamefont {C.}~\bibnamefont
  {Honerkamp}},\ }\href {https://doi.org/10.1143/PTP.105.1} {\bibfield
  {journal} {\bibinfo  {journal} {Prog. Theor. Phys.}\ }\textbf {\bibinfo
  {volume} {105}},\ \bibinfo {pages} {1} (\bibinfo {year} {2001})}\BibitemShut
  {NoStop}%
\bibitem [{\citenamefont {Kugler}\ and\ \citenamefont {von
  Delft}(2018{\natexlab{a}})}]{KuglerMfrgPRL2018}%
  \BibitemOpen
  \bibfield  {author} {\bibinfo {author} {\bibfnamefont {F.~B.}\ \bibnamefont
  {Kugler}}\ and\ \bibinfo {author} {\bibfnamefont {J.}~\bibnamefont {von
  Delft}},\ }\href {https://doi.org/10.1103/PhysRevLett.120.057403} {\bibfield
  {journal} {\bibinfo  {journal} {Phys. Rev. Lett.}\ }\textbf {\bibinfo
  {volume} {120}},\ \bibinfo {pages} {057403} (\bibinfo {year}
  {2018}{\natexlab{a}})}\BibitemShut {NoStop}%
\bibitem [{\citenamefont {Kugler}\ and\ \citenamefont {von
  Delft}(2018{\natexlab{b}})}]{KuglerGeneral2018}%
  \BibitemOpen
  \bibfield  {author} {\bibinfo {author} {\bibfnamefont {F.~B.}\ \bibnamefont
  {Kugler}}\ and\ \bibinfo {author} {\bibfnamefont {J.}~\bibnamefont {von
  Delft}},\ }\href {https://doi.org/10.1103/PhysRevB.97.035162} {\bibfield
  {journal} {\bibinfo  {journal} {Phys. Rev. B}\ }\textbf {\bibinfo {volume}
  {97}},\ \bibinfo {pages} {035162} (\bibinfo {year}
  {2018}{\natexlab{b}})}\BibitemShut {NoStop}%
\bibitem [{\citenamefont {Kugler}\ and\ \citenamefont {von
  Delft}(2018{\natexlab{c}})}]{KuglerParquet2018}%
  \BibitemOpen
  \bibfield  {author} {\bibinfo {author} {\bibfnamefont {F.~B.}\ \bibnamefont
  {Kugler}}\ and\ \bibinfo {author} {\bibfnamefont {J.}~\bibnamefont {von
  Delft}},\ }\href {https://doi.org/10.1088/1367-2630/aaf65f} {\bibfield
  {journal} {\bibinfo  {journal} {New J. Phys.}\ }\textbf {\bibinfo {volume}
  {20}},\ \bibinfo {pages} {123029} (\bibinfo {year}
  {2018}{\natexlab{c}})}\BibitemShut {NoStop}%
\bibitem [{\citenamefont {Chalupa-Gantner}\ \emph {et~al.}(2022)\citenamefont
  {Chalupa-Gantner}, \citenamefont {Kugler}, \citenamefont {Hille},
  \citenamefont {von Delft}, \citenamefont {Andergassen},\ and\ \citenamefont
  {Toschi}}]{ChalupaGantner2022}%
  \BibitemOpen
  \bibfield  {author} {\bibinfo {author} {\bibfnamefont {P.}~\bibnamefont
  {Chalupa-Gantner}}, \bibinfo {author} {\bibfnamefont {F.~B.}\ \bibnamefont
  {Kugler}}, \bibinfo {author} {\bibfnamefont {C.}~\bibnamefont {Hille}},
  \bibinfo {author} {\bibfnamefont {J.}~\bibnamefont {von Delft}}, \bibinfo
  {author} {\bibfnamefont {S.}~\bibnamefont {Andergassen}},\ and\ \bibinfo
  {author} {\bibfnamefont {A.}~\bibnamefont {Toschi}},\ }\href
  {https://doi.org/10.1103/PhysRevResearch.4.023050} {\bibfield  {journal}
  {\bibinfo  {journal} {Phys. Rev. Research}\ }\textbf {\bibinfo {volume}
  {4}},\ \bibinfo {pages} {023050} (\bibinfo {year} {2022})}\BibitemShut
  {NoStop}%
\bibitem [{\citenamefont {Khoromskij}(2011)}]{Khoromskij2011}%
  \BibitemOpen
  \bibfield  {author} {\bibinfo {author} {\bibfnamefont {B.~N.}\ \bibnamefont
  {Khoromskij}},\ }\href {https://doi.org/10.1007/s00365-011-9131-1} {\bibfield
   {journal} {\bibinfo  {journal} {Constr. Approx.}\ }\textbf {\bibinfo
  {volume} {34}},\ \bibinfo {pages} {257} (\bibinfo {year} {2011})}\BibitemShut
  {NoStop}%
\bibitem [{\citenamefont {Oseledets}(2011)}]{Oseledets2011}%
  \BibitemOpen
  \bibfield  {author} {\bibinfo {author} {\bibfnamefont {I.~V.}\ \bibnamefont
  {Oseledets}},\ }\href {https://doi.org/10.1137/090752286} {\bibfield
  {journal} {\bibinfo  {journal} {SIAM J. Sci. Comput.}\ }\textbf {\bibinfo
  {volume} {33}},\ \bibinfo {pages} {2295} (\bibinfo {year}
  {2011})}\BibitemShut {NoStop}%
\bibitem [{\citenamefont {N{\'u}{\~n}ez~Fern{\'a}ndez}\ \emph
  {et~al.}(2022)\citenamefont {N{\'u}{\~n}ez~Fern{\'a}ndez}, \citenamefont
  {Jeannin}, \citenamefont {Dumitrescu}, \citenamefont {Kloss}, \citenamefont
  {Kaye}, \citenamefont {Parcollet},\ and\ \citenamefont
  {Waintal}}]{NunezFernandez2022}%
  \BibitemOpen
  \bibfield  {author} {\bibinfo {author} {\bibfnamefont {Y.}~\bibnamefont
  {N{\'u}{\~n}ez~Fern{\'a}ndez}}, \bibinfo {author} {\bibfnamefont
  {M.}~\bibnamefont {Jeannin}}, \bibinfo {author} {\bibfnamefont {P.~T.}\
  \bibnamefont {Dumitrescu}}, \bibinfo {author} {\bibfnamefont
  {T.}~\bibnamefont {Kloss}}, \bibinfo {author} {\bibfnamefont
  {J.}~\bibnamefont {Kaye}}, \bibinfo {author} {\bibfnamefont {O.}~\bibnamefont
  {Parcollet}},\ and\ \bibinfo {author} {\bibfnamefont {X.}~\bibnamefont
  {Waintal}},\ }\href {https://doi.org/10.1103/PhysRevX.12.041018} {\bibfield
  {journal} {\bibinfo  {journal} {Phys. Rev. X}\ }\textbf {\bibinfo {volume}
  {12}},\ \bibinfo {pages} {041018} (\bibinfo {year} {2022})}\BibitemShut
  {NoStop}%
\bibitem [{\citenamefont {Ritter}\ \emph {et~al.}(2024)\citenamefont {Ritter},
  \citenamefont {N{\'u}{\~n}ez~Fern{\'a}ndez}, \citenamefont {Wallerberger},
  \citenamefont {von Delft}, \citenamefont {Shinaoka},\ and\ \citenamefont
  {Waintal}}]{Ritter2024}%
  \BibitemOpen
  \bibfield  {author} {\bibinfo {author} {\bibfnamefont {M.~K.}\ \bibnamefont
  {Ritter}}, \bibinfo {author} {\bibfnamefont {Y.}~\bibnamefont
  {N{\'u}{\~n}ez~Fern{\'a}ndez}}, \bibinfo {author} {\bibfnamefont
  {M.}~\bibnamefont {Wallerberger}}, \bibinfo {author} {\bibfnamefont
  {J.}~\bibnamefont {von Delft}}, \bibinfo {author} {\bibfnamefont
  {H.}~\bibnamefont {Shinaoka}},\ and\ \bibinfo {author} {\bibfnamefont
  {X.}~\bibnamefont {Waintal}},\ }\href
  {https://doi.org/10.1103/PhysRevLett.132.056501} {\bibfield  {journal}
  {\bibinfo  {journal} {Phys. Rev. Lett.}\ }\textbf {\bibinfo {volume} {132}},\
  \bibinfo {pages} {056501} (\bibinfo {year} {2024})}\BibitemShut {NoStop}%
\bibitem [{\citenamefont {Ishida}\ \emph {et~al.}(2025)\citenamefont {Ishida},
  \citenamefont {Okada}, \citenamefont {Hoshino},\ and\ \citenamefont
  {Shinaoka}}]{Ishida2025}%
  \BibitemOpen
  \bibfield  {author} {\bibinfo {author} {\bibfnamefont {H.}~\bibnamefont
  {Ishida}}, \bibinfo {author} {\bibfnamefont {N.}~\bibnamefont {Okada}},
  \bibinfo {author} {\bibfnamefont {S.}~\bibnamefont {Hoshino}},\ and\ \bibinfo
  {author} {\bibfnamefont {H.}~\bibnamefont {Shinaoka}},\ }\href
  {https://doi.org/10.1103/tkcp-p5br} {\bibfield  {journal} {\bibinfo
  {journal} {Phys. Rev. Lett.}\ }\textbf {\bibinfo {volume} {135}},\ \bibinfo
  {pages} {046502} (\bibinfo {year} {2025})}\BibitemShut {NoStop}%
\bibitem [{\citenamefont {Frankenbach}\ \emph {et~al.}(2025)\citenamefont
  {Frankenbach}, \citenamefont {Ritter}, \citenamefont {Pelz}, \citenamefont
  {Ritz}, \citenamefont {von Delft},\ and\ \citenamefont
  {Ge}}]{Frankenbach2025}%
  \BibitemOpen
  \bibfield  {author} {\bibinfo {author} {\bibfnamefont {M.}~\bibnamefont
  {Frankenbach}}, \bibinfo {author} {\bibfnamefont {M.~K.}\ \bibnamefont
  {Ritter}}, \bibinfo {author} {\bibfnamefont {M.}~\bibnamefont {Pelz}},
  \bibinfo {author} {\bibfnamefont {N.}~\bibnamefont {Ritz}}, \bibinfo {author}
  {\bibfnamefont {J.}~\bibnamefont {von Delft}},\ and\ \bibinfo {author}
  {\bibfnamefont {A.}~\bibnamefont {Ge}},\ }\href
  {https://doi.org/10.1103/jx7h-lsqk} {\bibfield  {journal} {\bibinfo
  {journal} {Phys. Rev. Research}\ }\textbf {\bibinfo {volume} {7}},\ \bibinfo
  {pages} {043032} (\bibinfo {year} {2025})}\BibitemShut {NoStop}%
\bibitem [{\citenamefont {Ge}\ \emph {et~al.}(2024)\citenamefont {Ge},
  \citenamefont {Ritz}, \citenamefont {Walter}, \citenamefont {Aguirre},
  \citenamefont {von Delft},\ and\ \citenamefont {Kugler}}]{Ge2024}%
  \BibitemOpen
  \bibfield  {author} {\bibinfo {author} {\bibfnamefont {A.}~\bibnamefont
  {Ge}}, \bibinfo {author} {\bibfnamefont {N.}~\bibnamefont {Ritz}}, \bibinfo
  {author} {\bibfnamefont {E.}~\bibnamefont {Walter}}, \bibinfo {author}
  {\bibfnamefont {S.}~\bibnamefont {Aguirre}}, \bibinfo {author} {\bibfnamefont
  {J.}~\bibnamefont {von Delft}},\ and\ \bibinfo {author} {\bibfnamefont
  {F.~B.}\ \bibnamefont {Kugler}},\ }\href
  {https://doi.org/10.1103/PhysRevB.109.115128} {\bibfield  {journal} {\bibinfo
   {journal} {Phys. Rev. B}\ }\textbf {\bibinfo {volume} {109}},\ \bibinfo
  {pages} {115128} (\bibinfo {year} {2024})}\BibitemShut {NoStop}%
\bibitem [{\citenamefont {Ge}\ \emph {et~al.}(2023)\citenamefont {Ge},
  \citenamefont {Ritz}, \citenamefont {Walter}, \citenamefont {Aguirre},
  \citenamefont {von Delft},\ and\ \citenamefont {Kugler}}]{Ge2023}%
  \BibitemOpen
  \bibfield  {author} {\bibinfo {author} {\bibfnamefont {A.}~\bibnamefont
  {Ge}}, \bibinfo {author} {\bibfnamefont {N.}~\bibnamefont {Ritz}}, \bibinfo
  {author} {\bibfnamefont {E.}~\bibnamefont {Walter}}, \bibinfo {author}
  {\bibfnamefont {S.}~\bibnamefont {Aguirre}}, \bibinfo {author} {\bibfnamefont
  {J.}~\bibnamefont {von Delft}},\ and\ \bibinfo {author} {\bibfnamefont
  {F.}~\bibnamefont {Kugler}},\ }\href {https://doi.org/10.57970/xeqvs-w3p49}
  {\bibinfo {title} {Data for: Real-frequency quantum field theory applied to
  the single-impurity {A}nderson model}},\ \bibinfo {howpublished} {{LMU}
  Munich, Faculty of Physics} (\bibinfo {year} {2023}),\ \bibinfo {note}
  {dataset}\BibitemShut {NoStop}%
\bibitem [{\citenamefont {Katanin}(2004)}]{Katanin2004}%
  \BibitemOpen
  \bibfield  {author} {\bibinfo {author} {\bibfnamefont {A.~A.}\ \bibnamefont
  {Katanin}},\ }\href {https://doi.org/10.1103/PhysRevB.70.115109} {\bibfield
  {journal} {\bibinfo  {journal} {Phys. Rev. B}\ }\textbf {\bibinfo {volume}
  {70}},\ \bibinfo {pages} {115109} (\bibinfo {year} {2004})}\BibitemShut
  {NoStop}%
\bibitem [{\citenamefont {Yamada}(1975)}]{Yamada1975a}%
  \BibitemOpen
  \bibfield  {author} {\bibinfo {author} {\bibfnamefont {K.}~\bibnamefont
  {Yamada}},\ }\href {https://doi.org/10.1143/PTP.53.970} {\bibfield  {journal}
  {\bibinfo  {journal} {Prog. Theor. Phys.}\ }\textbf {\bibinfo {volume}
  {53}},\ \bibinfo {pages} {970} (\bibinfo {year} {1975})}\BibitemShut
  {NoStop}%
\bibitem [{\citenamefont {Hewson}(1993)}]{Hewson1993}%
  \BibitemOpen
  \bibfield  {author} {\bibinfo {author} {\bibfnamefont {A.~C.}\ \bibnamefont
  {Hewson}},\ }\href {https://doi.org/10.1017/CBO9780511470752} {\emph
  {\bibinfo {title} {The {K}ondo Problem to Heavy Fermions}}}\ (\bibinfo
  {publisher} {Cambridge University Press},\ \bibinfo {address} {Cambridge},\
  \bibinfo {year} {1993})\BibitemShut {NoStop}%
\bibitem [{\citenamefont {Hewson}(2001)}]{Hewson2001}%
  \BibitemOpen
  \bibfield  {author} {\bibinfo {author} {\bibfnamefont {A.~C.}\ \bibnamefont
  {Hewson}},\ }\href {https://doi.org/10.1088/0953-8984/13/44/314} {\bibfield
  {journal} {\bibinfo  {journal} {J. Phys.: Condens. Matter}\ }\textbf
  {\bibinfo {volume} {13}},\ \bibinfo {pages} {10011} (\bibinfo {year}
  {2001})}\BibitemShut {NoStop}%
\bibitem [{\citenamefont {Vidberg}\ and\ \citenamefont
  {Serene}(1977)}]{Vidberg1977}%
  \BibitemOpen
  \bibfield  {author} {\bibinfo {author} {\bibfnamefont {H.~J.}\ \bibnamefont
  {Vidberg}}\ and\ \bibinfo {author} {\bibfnamefont {J.~W.}\ \bibnamefont
  {Serene}},\ }\href {https://doi.org/10.1007/BF00655090} {\bibfield  {journal}
  {\bibinfo  {journal} {J. Low Temp. Phys.}\ }\textbf {\bibinfo {volume}
  {29}},\ \bibinfo {pages} {179} (\bibinfo {year} {1977})}\BibitemShut
  {NoStop}%
\bibitem [{\citenamefont {Kanamori}(1963)}]{Kanamori1963}%
  \BibitemOpen
  \bibfield  {author} {\bibinfo {author} {\bibfnamefont {J.}~\bibnamefont
  {Kanamori}},\ }\href {https://doi.org/10.1143/PTP.30.275} {\bibfield
  {journal} {\bibinfo  {journal} {Prog. Theor. Phys.}\ }\textbf {\bibinfo
  {volume} {30}},\ \bibinfo {pages} {275} (\bibinfo {year} {1963})}\BibitemShut
  {NoStop}%
\bibitem [{\citenamefont {de' Medici}\ \emph {et~al.}(2011)\citenamefont {de'
  Medici}, \citenamefont {Mravlje},\ and\ \citenamefont
  {Georges}}]{deMedici2011}%
  \BibitemOpen
  \bibfield  {author} {\bibinfo {author} {\bibfnamefont {L.}~\bibnamefont {de'
  Medici}}, \bibinfo {author} {\bibfnamefont {J.}~\bibnamefont {Mravlje}},\
  and\ \bibinfo {author} {\bibfnamefont {A.}~\bibnamefont {Georges}},\ }\href
  {https://doi.org/10.1103/PhysRevLett.107.256401} {\bibfield  {journal}
  {\bibinfo  {journal} {Phys. Rev. Lett.}\ }\textbf {\bibinfo {volume} {107}},\
  \bibinfo {pages} {256401} (\bibinfo {year} {2011})}\BibitemShut {NoStop}%
\bibitem [{\citenamefont {Stadler}\ \emph {et~al.}(2015)\citenamefont
  {Stadler}, \citenamefont {Yin}, \citenamefont {von Delft}, \citenamefont
  {Kotliar},\ and\ \citenamefont {Weichselbaum}}]{Stadler2015}%
  \BibitemOpen
  \bibfield  {author} {\bibinfo {author} {\bibfnamefont {K.~M.}\ \bibnamefont
  {Stadler}}, \bibinfo {author} {\bibfnamefont {Z.~P.}\ \bibnamefont {Yin}},
  \bibinfo {author} {\bibfnamefont {J.}~\bibnamefont {von Delft}}, \bibinfo
  {author} {\bibfnamefont {G.}~\bibnamefont {Kotliar}},\ and\ \bibinfo {author}
  {\bibfnamefont {A.}~\bibnamefont {Weichselbaum}},\ }\href
  {https://doi.org/10.1103/PhysRevLett.115.136401} {\bibfield  {journal}
  {\bibinfo  {journal} {Phys. Rev. Lett.}\ }\textbf {\bibinfo {volume} {115}},\
  \bibinfo {pages} {136401} (\bibinfo {year} {2015})}\BibitemShut {NoStop}%
\bibitem [{\citenamefont {{\v{S}}mejkal}\ \emph {et~al.}(2022)\citenamefont
  {{\v{S}}mejkal}, \citenamefont {Sinova},\ and\ \citenamefont
  {Jungwirth}}]{Smejkal2022PRX}%
  \BibitemOpen
  \bibfield  {author} {\bibinfo {author} {\bibfnamefont {L.}~\bibnamefont
  {{\v{S}}mejkal}}, \bibinfo {author} {\bibfnamefont {J.}~\bibnamefont
  {Sinova}},\ and\ \bibinfo {author} {\bibfnamefont {T.}~\bibnamefont
  {Jungwirth}},\ }\href {https://doi.org/10.1103/PhysRevX.12.031042} {\bibfield
   {journal} {\bibinfo  {journal} {Phys. Rev. X}\ }\textbf {\bibinfo {volume}
  {12}},\ \bibinfo {pages} {031042} (\bibinfo {year} {2022})}\BibitemShut
  {NoStop}%
\bibitem [{\citenamefont {Krempask{\'y}}\ \emph {et~al.}(2024)\citenamefont
  {Krempask{\'y}}, \citenamefont {{\v{S}}mejkal}, \citenamefont {D'Souza},
  \citenamefont {Hajlaoui}, \citenamefont {Springholz}, \citenamefont
  {Uhl{\'i}{\v{r}}ov{\'a}}, \citenamefont {Alarab}, \citenamefont
  {Constantinou}, \citenamefont {Strocov}, \citenamefont {Usanov},
  \citenamefont {Pudelko}, \citenamefont {Gonz{\'a}lez-Hern{\'a}ndez},
  \citenamefont {Birk~Hellenes}, \citenamefont {Jansa}, \citenamefont
  {Reichlov{\'a}}, \citenamefont {{\v{S}}ob{\'a}{\v{n}}}, \citenamefont
  {Gonzalez~Betancourt}, \citenamefont {Wadley}, \citenamefont {Sinova},
  \citenamefont {Kriegner}, \citenamefont {Min{\'a}r}, \citenamefont {Dil},\
  and\ \citenamefont {Jungwirth}}]{Krempasky2024}%
  \BibitemOpen
  \bibfield  {author} {\bibinfo {author} {\bibfnamefont {J.}~\bibnamefont
  {Krempask{\'y}}}, \bibinfo {author} {\bibfnamefont {L.}~\bibnamefont
  {{\v{S}}mejkal}}, \bibinfo {author} {\bibfnamefont {S.~W.}\ \bibnamefont
  {D'Souza}}, \bibinfo {author} {\bibfnamefont {M.}~\bibnamefont {Hajlaoui}},
  \bibinfo {author} {\bibfnamefont {G.}~\bibnamefont {Springholz}}, \bibinfo
  {author} {\bibfnamefont {K.}~\bibnamefont {Uhl{\'i}{\v{r}}ov{\'a}}}, \bibinfo
  {author} {\bibfnamefont {F.}~\bibnamefont {Alarab}}, \bibinfo {author}
  {\bibfnamefont {P.~C.}\ \bibnamefont {Constantinou}}, \bibinfo {author}
  {\bibfnamefont {V.}~\bibnamefont {Strocov}}, \bibinfo {author} {\bibfnamefont
  {D.}~\bibnamefont {Usanov}}, \bibinfo {author} {\bibfnamefont {W.~R.}\
  \bibnamefont {Pudelko}}, \bibinfo {author} {\bibfnamefont {R.}~\bibnamefont
  {Gonz{\'a}lez-Hern{\'a}ndez}}, \bibinfo {author} {\bibfnamefont
  {A.}~\bibnamefont {Birk~Hellenes}}, \bibinfo {author} {\bibfnamefont
  {Z.}~\bibnamefont {Jansa}}, \bibinfo {author} {\bibfnamefont
  {H.}~\bibnamefont {Reichlov{\'a}}}, \bibinfo {author} {\bibfnamefont
  {Z.}~\bibnamefont {{\v{S}}ob{\'a}{\v{n}}}}, \bibinfo {author} {\bibfnamefont
  {R.~D.}\ \bibnamefont {Gonzalez~Betancourt}}, \bibinfo {author}
  {\bibfnamefont {P.}~\bibnamefont {Wadley}}, \bibinfo {author} {\bibfnamefont
  {J.}~\bibnamefont {Sinova}}, \bibinfo {author} {\bibfnamefont
  {D.}~\bibnamefont {Kriegner}}, \bibinfo {author} {\bibfnamefont
  {J.}~\bibnamefont {Min{\'a}r}}, \bibinfo {author} {\bibfnamefont {J.~H.}\
  \bibnamefont {Dil}},\ and\ \bibinfo {author} {\bibfnamefont {T.}~\bibnamefont
  {Jungwirth}},\ }\href {https://doi.org/10.1038/s41586-023-06907-7} {\bibfield
   {journal} {\bibinfo  {journal} {Nature}\ }\textbf {\bibinfo {volume}
  {626}},\ \bibinfo {pages} {517} (\bibinfo {year} {2024})}\BibitemShut
  {NoStop}%
\bibitem [{\citenamefont {Vila}\ \emph {et~al.}(2025)\citenamefont {Vila},
  \citenamefont {Sunko},\ and\ \citenamefont {Moore}}]{Vila2025}%
  \BibitemOpen
  \bibfield  {author} {\bibinfo {author} {\bibfnamefont {M.}~\bibnamefont
  {Vila}}, \bibinfo {author} {\bibfnamefont {V.}~\bibnamefont {Sunko}},\ and\
  \bibinfo {author} {\bibfnamefont {J.~E.}\ \bibnamefont {Moore}},\ }\href
  {https://doi.org/10.1103/bzzy-ngcs} {\bibfield  {journal} {\bibinfo
  {journal} {Phys. Rev. B}\ }\textbf {\bibinfo {volume} {112}},\ \bibinfo
  {pages} {L020401} (\bibinfo {year} {2025})}\BibitemShut {NoStop}%
\bibitem [{\citenamefont {Diniz}\ and\ \citenamefont
  {Vernek}(2024)}]{Diniz2024}%
  \BibitemOpen
  \bibfield  {author} {\bibinfo {author} {\bibfnamefont {G.~S.}\ \bibnamefont
  {Diniz}}\ and\ \bibinfo {author} {\bibfnamefont {E.}~\bibnamefont {Vernek}},\
  }\href {https://doi.org/10.1103/PhysRevB.109.155127} {\bibfield  {journal}
  {\bibinfo  {journal} {Phys. Rev. B}\ }\textbf {\bibinfo {volume} {109}},\
  \bibinfo {pages} {155127} (\bibinfo {year} {2024})}\BibitemShut {NoStop}%
\bibitem [{\citenamefont {Aoyama}\ and\ \citenamefont
  {Ohgushi}(2024)}]{Aoyama2024}%
  \BibitemOpen
  \bibfield  {author} {\bibinfo {author} {\bibfnamefont {T.}~\bibnamefont
  {Aoyama}}\ and\ \bibinfo {author} {\bibfnamefont {K.}~\bibnamefont
  {Ohgushi}},\ }\href {https://doi.org/10.1103/PhysRevMaterials.8.L041402}
  {\bibfield  {journal} {\bibinfo  {journal} {Phys. Rev. Mater.}\ }\textbf
  {\bibinfo {volume} {8}},\ \bibinfo {pages} {L041402} (\bibinfo {year}
  {2024})}\BibitemShut {NoStop}%
\bibitem [{\citenamefont {Crosswhite}\ and\ \citenamefont
  {Bacon}(2008)}]{Crosswhite2008}%
  \BibitemOpen
  \bibfield  {author} {\bibinfo {author} {\bibfnamefont {G.~M.}\ \bibnamefont
  {Crosswhite}}\ and\ \bibinfo {author} {\bibfnamefont {D.}~\bibnamefont
  {Bacon}},\ }\href {https://doi.org/10.1103/PhysRevA.78.012356} {\bibfield
  {journal} {\bibinfo  {journal} {Phys. Rev. A}\ }\textbf {\bibinfo {volume}
  {78}},\ \bibinfo {pages} {012356} (\bibinfo {year} {2008})}\BibitemShut
  {NoStop}%
\bibitem [{\citenamefont {Kazeev}\ and\ \citenamefont
  {Khoromskij}(2012)}]{Kazeev2012}%
  \BibitemOpen
  \bibfield  {author} {\bibinfo {author} {\bibfnamefont {V.~A.}\ \bibnamefont
  {Kazeev}}\ and\ \bibinfo {author} {\bibfnamefont {B.~N.}\ \bibnamefont
  {Khoromskij}},\ }\href {https://doi.org/10.1137/100820479} {\bibfield
  {journal} {\bibinfo  {journal} {SIAM J. Matrix Anal. Appl.}\ }\textbf
  {\bibinfo {volume} {33}},\ \bibinfo {pages} {742} (\bibinfo {year}
  {2012})}\BibitemShut {NoStop}%
\end{thebibliography}%

\clearpage
\onecolumngrid
\begin{center}
{\large\bfseries Supplementary Information\par}
\vspace{4pt}
{Quantics tensor trains for multiloop functional renormalization group in multiorbital impurity models\par}
\vspace{4pt}
{Hyun-Yong Lee and Naoki Kawashima\par}
\end{center}
\vspace{8pt}
\twocolumngrid
\setcounter{equation}{0}
\setcounter{figure}{0}
\setcounter{table}{0}
\setcounter{paperalgorithm}{0}
\renewcommand{\theequation}{S\arabic{equation}}
\renewcommand{\thefigure}{S\arabic{figure}}
\renewcommand{\thetable}{S\arabic{table}}

\section*{Supplementary Note 1. Algebraic identities}

The following identities supplement the tensor definitions in the main Methods. {We use the flavor bases of main-text Eqs.~(4) and (16), the channel coordinates of Eq.~(15), and the carry tensors of Eq.~(20). The particle--particle, direct particle--hole, and crossed particle--hole channels are denoted by $\mathrm{pp}$, $\mathrm{ph}$, and $\phbar$, respectively.} \rev{Frequency sums run over Matsubara indices; the internal fermionic Matsubara frequency is $\nu_k$.} The frequency transformations, contraction sequence, truncation rules, and numerical parameters used to obtain the reported results are specified there. The tables below collect the coefficients obtained from those definitions and do not introduce additional approximations.

\subsection*{Flavor factorization}

For the invariant spin and orbital tensors defined in the main text, contraction over the internal flavors produces
{
\begin{equation}
 C^{(r)}_{cab}=\sum_\alpha O^{(r)}_{c\alpha}
 L^{(r)}_{\alpha a}L^{(r)}_{\alpha b}.
\end{equation}
}
{Tables~\ref{meth:flavor-one} and \ref{meth:flavor-two} give exact factorizations for the one-orbital and degenerate two-orbital models. The first-input index $a$, second-input index $b$, and output index $c$ run over $(D,X)$ or $(1,\ldots,6)$, respectively. Each row corresponds to one value of $\alpha$; input and output vectors follow these basis orders. The coefficients follow from contractions of the Kronecker tensors. Interaction dependence enters through the vertex amplitudes, not these tables. The six-component bare vertex is}
\begin{equation}
 \boldsymbol\Gamma_0=U(1,-1,0,0,0,0)
 +J(-2,2,1,1,-1,-1),
\end{equation}
for the Kanamori interaction\cite{Kanamori1963,Georges2013}.

\begin{table}[tb]
\centering\small
\caption{Exact flavor factors for the single-orbital model {in the ordered spin basis $(D,X)$}.}
\label{meth:flavor-one}
\begin{tabular}{ccc}\hline
$r$&{$L_{\alpha a}$}&{$O_{c\alpha}$}\\\hline
$\mathrm{ph}$&$(2,1)$&$(1/2,0)$\\
&$(0,1)$&$(-1/2,1)$\\
$\mathrm{pp}$&$(1,1)$&$(1/2,1/2)$\\
&$(1,-1)$&$(1/2,-1/2)$\\
$\phbar$&$(1,0)$&$(1,-1/2)$\\
&$(1,2)$&$(0,1/2)$\\\hline
\end{tabular}
\end{table}
\begin{table*}[tb]
\centering\small
\caption{Exact flavor factors in the two-orbital basis of the main-text Methods.}
\label{meth:flavor-two}
\begin{tabular}{ccc}\hline
$r$&{$L_{\alpha a}$}&{$O_{c\alpha}$}\\\hline
$\mathrm{ph}$&$(0,1,0,0,0,1)$&$(1/4,1/2,-1/4,-1/4,-1/2,1/2)$\\
&$(4,1,2,2,2,1)$&$(1/4,0,0,0,0,0)$\\
&$(0,1,0,0,0,-1)$&$(0,1/2,-1/4,1/4,0,-1/2)$\\
&$(0,1,2,-2,0,-1)$&$(0,0,1/4,-1/4,0,0)$\\
&$(0,1,2,2,0,1)$&$(-1/4,0,1/4,1/4,0,0)$\\
&$(0,1,0,0,2,1)$&$(-1/4,0,0,0,1/2,0)$\\\hline
$\mathrm{pp}$&$(1,1,1,0,1,0)$&$(1/4,1/4,1/4,-1/4,1/4,-1/4)$\\
&$(1,1,-1,0,-1,0)$&$(1/4,1/4,-1/4,0,-1/4,0)$\\
&$(1,-1,1,0,-1,0)$&$(1/4,-1/4,1/4,-1/4,-1/4,1/4)$\\
&$(1,-1,-1,0,1,0)$&$(1/4,-1/4,-1/4,0,1/4,0)$\\
&$(1,1,1,2,1,2)$&$(0,0,0,1/4,0,1/4)$\\
&$(1,-1,1,2,-1,-2)$&$(0,0,0,1/4,0,-1/4)$\\\hline
$\phbar$&$(1,0,0,1,0,0)$&$(1/2,1/4,-1/2,1/2,-1/4,-1/4)$\\
&$(1,0,0,-1,0,0)$&$(1/2,0,0,-1/2,-1/4,1/4)$\\
&$(1,4,2,1,2,2)$&$(0,1/4,0,0,0,0)$\\
&$(1,0,0,-1,2,-2)$&$(0,0,0,0,1/4,-1/4)$\\
&$(1,0,2,1,0,0)$&$(0,-1/4,1/2,0,0,0)$\\
&$(1,0,0,1,2,2)$&$(0,-1/4,0,0,1/4,1/4)$\\\hline
\end{tabular}
\end{table*}

\subsection*{Integer frequency identities}

A channel transformation matches the same four external frequency labels in two parametrizations. Fermionic reversal is $\nu_n\mapsto\nu_{-n-1}$, so the constant integer shifts in Table~\ref{meth:frequency-maps} cannot be omitted. The entries give source coordinates $(m_s,n_s,n'_s)$ as functions of the four target coordinates $(m,n,n',k)$. They combine the external leg assignment with the internal-frequency substitution of each bubble factor.

{The factors $A$ and $B$ are the first and second inputs of $\Bcal_r[A,P,B]$, respectively, not their geometric positions in the $\mathrm{ph}$ diagram. The target labels represent $(\Omega_m,\nu_n,\nu_{n'},\nu_k)$. In particular, the second $\mathrm{pp}$ input contains $-m-k-1$ because $-\Omega_m-\nu_k=\nu_{-m-k-1}$.}

The carry construction follows directly from binary arithmetic. At site $j$, \rev{where $\boldsymbol x_j$ and $\boldsymbol y_j$ collect the $j$th binary digits of the target and source offset codes,} its defining equation is
\begin{equation}
 2\boldsymbol c_{j+1}+\boldsymbol y_j
 =\boldsymbol c_j+\boldsymbol\beta_j+\mathbf M\boldsymbol x_j.
\end{equation}
Multiplying by $2^j$ and summing telescopes the carries:
\begin{equation}
 \sum_j2^j\boldsymbol y_j
 =\mathbf M\sum_j2^j\boldsymbol x_j
 +\sum_j2^j\boldsymbol\beta_j+\boldsymbol c_0
 -2^R\boldsymbol c_R.
\end{equation}
The prescribed left boundary and the remaining right-boundary propagation give precisely the affine relation between offset codes, with no periodic identification. An out-of-domain source coordinate has no accepted boundary carry and contributes zero. \rev{Since the digits are $0$ or $1$ and $|\beta_{\mu j}|\le1$, the carries remain bounded integers, as stated in the main Methods.} This argument concerns the operator before tensor rounding; it does not assign zero numerical error to a compressed transformed tensor.

The central restriction from depth $R+1$ to $R$ is another binary identity. For $p=2^{R-1}$, the two encodings of a physical label satisfy
$x_{R+1}=x_R+p$. The low $R-1$ bits are unchanged. If the last bit of $x_R$ is $b$, the two last bits of $x_{R+1}$ are $(1-b,b)$. Contracting the final two cores with this selection yields the restriction tensor in the main Methods, separately for every frequency coordinate.

\begin{table*}[tb]
\centering\small
\caption{Source frequency coordinates for each channel and vertex factor.}
\label{meth:frequency-maps}
\begin{tabular}{ccccc}\hline
Output channel&Factor&$\mathrm{ph}$ source&$\mathrm{pp}$ source&$\phbar$ source\\\hline
$\mathrm{ph}$&{first input $A$}&$(m,k,n')$&$(-m-n'-k-1,k,n')$&$(n'-k,k,m+k)$\\
&{second input $B$}&$(m,n,k)$&$(-m-n-k-1,n,k)$&$(k-n,n,m+n)$\\
$\mathrm{pp}$&{first input $A$}&$(-m-n'-k-1,k,n')$&$(m,k,n')$&$(n'-k,k,-m-n'-1)$\\
&{second input $B$}&$(k-n,n,-m-k-1)$&$(m,n,-m-k-1)$&$(-m-n-k-1,n,k)$\\
$\phbar$&{first input $A$}&$(n'-k,k,m+k)$&$(-m-n'-k-1,k,m+k)$&$(m,k,n')$\\
&{second input $B$}&$(k-n,n,m+n)$&$(-m-n-k-1,n,m+n)$&$(m,n,k)$\\\hline
\end{tabular}
\end{table*}

\section*{Supplementary Note 2. Noninteracting and bare-vertex limits}

The response normalization can be fixed before an interacting reference is used. Define the single-flavor dressed bubble as
$\chi_0=-T\sum_\nu G(\ii\nu)G(\ii\nu+\ii\Omega)${$=-\Pi_0$ in main-text Eq.~(23)}.
For a spin-symmetric single orbital and a vanishing vertex,
\begin{equation}
 \chi_{\mathrm{charge}}=2\chi_0,\qquad
 {\chi_{S^z}}=\frac{\chi_0}{2},
 \qquad {S^z}=\frac{n_\uparrow-n_\downarrow}{2}.
\end{equation}
Retaining only the bare interaction gives
$\Gamma_{\uparrow\uparrow}=0$ and
$\Gamma_{\uparrow\downarrow}=U$ in the response contraction. Its two frequency sums factor independently:
\begin{equation}
 \chi_{\mathrm{charge}}=2\chi_0-2U\chi_0^2,\qquad
 {\chi_{S^z}}=\frac{\chi_0+U\chi_0^2}{2}.
\end{equation}
These identities constrain the relative channel sign, both temperature factors, and the spin normalization on the specified finite grid. They do not imply that a bare-vertex expression is adequate at finite interaction. The independently evaluated finite-grid sums reproduce them to machine precision.

\section*{Supplementary Note 3. Additional parquet comparisons}

\begin{figure}[t]
\centering
\includegraphics[width=\columnwidth]{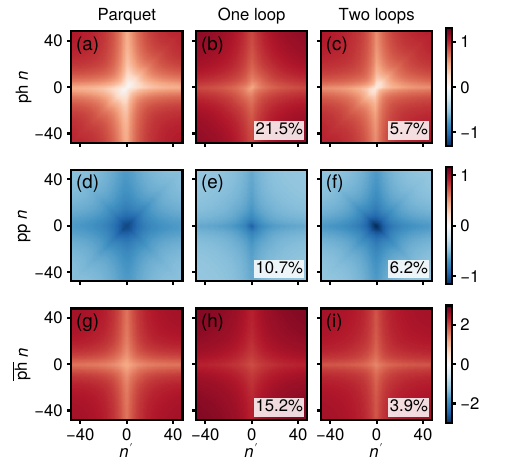}
\caption{Parquet, one-loop, and two-loop vertices (left to right) near $u=1$, $T/U=10^{-2}$, and $R=8$. {Rows show the $\mathrm{ph}$, $\mathrm{pp}$, and $\phbar$ channels}. Colors give {$\operatorname{Re}\gamma_{r,D}/U$} at zero bosonic transfer. Percentages use complete-plane deviations; central frequency windows are displayed.}
\label{sup:all-planes}
\end{figure}

{Figure~\ref{sup:all-planes} compares parquet, one-loop, and two-loop results in all three channels, using the Matsubara PA reference data\cite{Ge2024,Ge2023}.} Together the three channels yield the combined relative plane deviation reported in the main text. All complete planes enter with their actual amplitude, rather than equal channel weights. The displayed central regions are only visualization windows.

\section*{Supplementary Note 4. Second-order interpretation of the self-energy asymmetry}
\label{amsup:pt2-interpretation}

Second-order perturbation theory relates the self-energy asymmetry to competing Kanamori scattering processes. This derivation interprets the one-particle result; it does not replace the full frequency-dependent calculation. For noninteracting propagators $p=G_+^0(\tau)$ and $m=G_-^0(\tau)$ at the same regulator, particle--hole symmetry gives $G_s^0(-\tau)=-G_s^0(\tau)$. The bare Kanamori vertex therefore yields
\begin{align}
 \Sigma_+^{(2)}(\tau)&=\mathcal A p^3+\mathcal B p m^2,\nonumber\\
 \Sigma_-^{(2)}(\tau)&=\mathcal A m^3+\mathcal B m p^2,\nonumber\\
 \mathcal A&=(U-2J)^2,\qquad
 \mathcal B=U^2+(U-3J)^2+2J^2.
 \label{amsup:pt2-flavors}
\end{align}
The two $J^2$ terms in $\mathcal B$ are the spin-flip and pair-hopping contributions. Introducing $g=(p+m)/2$ and $d=(p-m)/2$ gives
\begin{align}
 \Sigma_+^{(2)}(\tau)-\Sigma_-^{(2)}(\tau)
 &=2d(3\mathcal A-\mathcal B)g^2\nonumber\\
 &\quad+2d^3(\mathcal A+\mathcal B),\nonumber\\
 3\mathcal A-\mathcal B&=U^2-6UJ+J^2.
 \label{amsup:pt2-asymmetry}
\end{align}
The term linear in bath asymmetry vanishes at $J/U=3-2\sqrt2$. At finite bath asymmetry, however, the cubic term in $d$ remains at the same interaction order and need not vanish there. Its frequency dependence can also move the zero differently for different averaging intervals. These are imaginary-time identities for the full convolution; restricting internal frequency domains adds finite-range dependence. Consequently, the difference between the {calculated zero} and $3-2\sqrt2$ is not by itself a quantitative measure of higher-order correlations. Such a comparison requires a second-order reference at the same bath asymmetry, temperature, regulator, and specified frequency domain.

\section*{Supplementary Note 5. Frequency dependence and additional orbital responses}
\label{amsup:additional-am-results}

\begin{figure}[!b]
\centering
\includegraphics[width=\columnwidth]{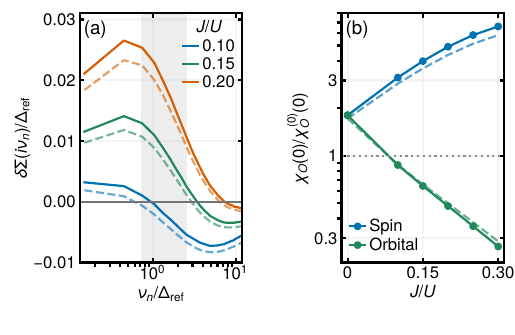}
\caption{Supporting altermagnetic results. (a) Self-energy difference at representative Hund couplings; shading indicates the main-text averaging interval. (b) Spin and orbital susceptibilities normalized by their noninteracting values. Solid and dashed curves denote $R=8$ and $7$, respectively. All results use the same one-loop endpoint.}
\label{sup:am-support}
\end{figure}

\begin{figure*}[t]
\centering
\includegraphics[width=\textwidth]{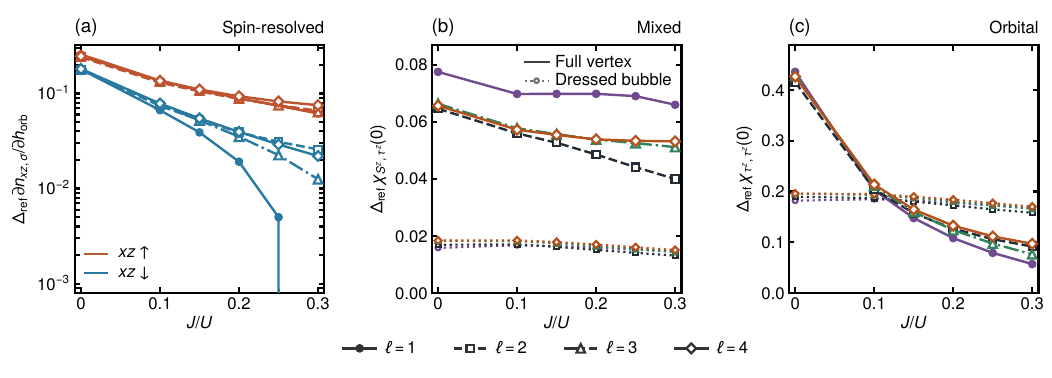}
\caption{Components of the orbital-field response at $R=10$, $T/U=10^{-2}$, and $\Lambda\simeq0.5\Delta_{\mathrm{ref}}$, at one to four loops. The one-loop $xz\downarrow$ response is negative at $J/U=0.30$ and lies outside the logarithmic axis of (a). Markers and line styles denote loop order; colors denote spin in (a) and loop order in (b,c), as in main-text Fig.~5. (a) Spin-resolved $xz$ occupancy responses. (b) Mixed susceptibility $\chi_{S^z,\tau^z}$. (c) Orbital susceptibility $\chi_{\tau^z,\tau^z}$. Full and dressed-bubble responses in (b,c) use the same propagators at each loop order. Hund coupling suppresses the orbital response more strongly than the mixed response.}
\label{sup:am-response-components}
\end{figure*}

The self-energy difference varies across the Matsubara domain, so its average and its lowest-frequency value are distinct observables. The {one-loop} profiles in Fig.~\ref{sup:am-support}(a) at $J/U=0.10$, $0.15$, and $0.20$ exhibit the frequency dependence underlying the main-text average over $n=2,\ldots,8$. {At one loop, this average changes sign between $J/U=0.10$ and $0.15$ at both $R=7$ and $8$. The loop-order comparison in main-text Fig.~5(c) instead holds the frequency range fixed at $R=10$.} A zero of this average does not require the complete self-energy difference to vanish. In particular, its sign cannot be interpreted as a frequency-independent shift of the host bands or as an independently determined effective-mass anisotropy.

The conventional spin and orbital responses in Fig.~\ref{sup:am-support}(b) provide a complementary reference for the Hund-coupling dependence. {The matrices $\mathbf O_{S^z}=\boldsymbol\sigma^z\otimes I_2/2$ and $\mathbf O_{\tau^z}=I_2\otimes\boldsymbol\tau^z/2$ use spin and orbital Pauli matrices, respectively (main Methods).} At $J=0$, the two responses coincide. Increasing Hund coupling enhances the spin susceptibility and suppresses the orbital susceptibility at both $R=7$ and $8$. The plotted ratios use the noninteracting susceptibility at the same bath, temperature, regulator, and frequency range. They do not use the dressed-bubble normalization of the separate degenerate two-orbital calculation.

The numerical sensitivity is particularly important for differences between responses. In the one-loop supplementary calculations, the total spin--orbital anisotropy {$\eta=\chi_{O_z,O_z}(0)-\chi_{O_x,O_x}(0)$, for $\hat O_\alpha=d^\dagger(\boldsymbol\sigma^\alpha\otimes\boldsymbol\tau^z/4)d$,} remains negative for all calculated Hund couplings at both frequency ranges, but its magnitude changes more strongly than either constituent susceptibility.

In the one-loop calculations at $R=7$ and $8$, the charge response exhibits a separate limitation: its static value is smaller than its value at the first nonzero bosonic Matsubara frequency, contrary to the expected equilibrium spectral ordering. This behavior enters through the vertex contribution, whereas the corresponding noninteracting and dressed-bubble values have the expected ordering. It is therefore recorded as an unresolved limitation of the approximate response and is not used as physical evidence; this decomposition does not establish that implementation errors are impossible.

\section*{Supplementary Note 6. Spin-resolved and absolute orbital-field responses}
\label{amsup:orbital-response}

The orbital-field spin polarization in the compensated altermagnetic environment follows from symmetry relations between the four densities. The spin-independent perturbation $H'=-h_{\mathrm{orb}}\tau^z$ lowers the $xz$ level and raises the $yz$ level for positive $h_{\mathrm{orb}}$. Using the main-text operators and static susceptibilities at zero orbital field, where the impurity has no magnetization, combined orbital interchange and spin reversal give
\begin{align}
 \frac{\partial\langle n_{xz\uparrow}\rangle}{\partial h_{\mathrm{orb}}}
 &=\frac{\chi_{\tau^z,\tau^z}+\chi_{S^z,\tau^z}}{2},\nonumber\\
 \frac{\partial\langle n_{xz\downarrow}\rangle}{\partial h_{\mathrm{orb}}}
 &=\frac{\chi_{\tau^z,\tau^z}-\chi_{S^z,\tau^z}}{2},\nonumber\\
 \frac{\partial\langle n_{yz\uparrow}\rangle}{\partial h_{\mathrm{orb}}}
 &=-\frac{\partial\langle n_{xz\downarrow}\rangle}{\partial h_{\mathrm{orb}}},\nonumber\\
 \frac{\partial\langle n_{yz\downarrow}\rangle}{\partial h_{\mathrm{orb}}}
 &=-\frac{\partial\langle n_{xz\uparrow}\rangle}{\partial h_{\mathrm{orb}}}.
 \label{amsup:density-decomposition}
\end{align}
The total charge response to this field vanishes. The sum and difference of the two $xz$ responses equal $\chi_{\tau^z,\tau^z}$ and $\chi_{S^z,\tau^z}$, respectively. The difference divided by the sum therefore measures the spin polarization of the induced $xz$ occupancy change, while the $yz$ orbital has the opposite occupancy changes with interchanged spin labels.

Electronic correlations change the spin composition of the altermagnetic orbital response. At two to four loops, both $xz$ responses in Fig.~\ref{sup:am-response-components}(a) remain positive, so their difference divided by their sum measures the polarization of occupancy changes in the same direction. Increasing Hund coupling suppresses both responses, but $xz\downarrow$ decreases more rapidly than $xz\uparrow$: at four loops, the $xz\uparrow$ response decreases by a factor of $3.3$ and the $xz\downarrow$ response by a factor of $8.2$ between $J=0$ and $J/U=0.30$. The trend is the same at two and three loops, whereas at one loop the $xz\downarrow$ response changes sign at $J/U=0.30$; the magnitude of the $xz\downarrow$ suppression at strong Hund coupling depends on loop order (Supplementary Note~8).

The absolute responses distinguish vertex enhancement at fixed Hund coupling from the variation with Hund coupling. Figure~\ref{sup:am-response-components}(b) shows that the full mixed susceptibility exceeds its dressed-bubble counterpart, yet decreases overall as Hund coupling grows. The orbital susceptibility decreases more strongly [Fig.~\ref{sup:am-response-components}(c)] and falls below its dressed bubble for $J/U\ge0.15$, producing the increasing polarization shown in the main text. Thus the vertex enhances the absolute mixed response relative to the same dressed propagators, but increasing Hund coupling reduces the magnitude of the induced magnetization at a fixed weak orbital field. These dressed-bubble comparisons are distinct from the noninteracting normalization in Fig.~\ref{sup:am-support}(b).

\section*{Supplementary Note 7. Orbital-field response at lower regulator cutoffs}
\label{amsup:cutoff-response}

\begin{figure}[t]
\centering
\includegraphics[width=\columnwidth]{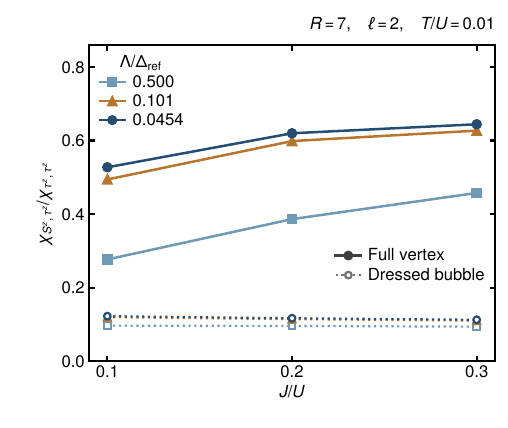}
\caption{Two-loop orbital-field spin polarization at three regulator cutoffs, with $R=7$ and $T/U=10^{-2}$. Full and dressed-bubble responses are compared at each cutoff. The increase with Hund coupling persists as the cutoff is lowered. Lines connect calculated points without fitting or extrapolation.}
\label{sup:am-response-cutoff}
\end{figure}

The increase of spin selectivity with Hund coupling persists at lower regulator cutoffs in the two-loop calculation. Figure~\ref{sup:am-response-cutoff} compares three cutoffs at $R=7$ and $T/U=10^{-2}$; the first cutoff is the endpoint used for the $R=10$ results in the main text. Lowering the cutoff increases the full-vertex polarization at each calculated Hund coupling, while its dependence on Hund coupling becomes less steep. The dressed-bubble values remain well below the full response at the corresponding cutoffs.

Both spin-resolved $xz$ responses remain positive throughout this comparison. It establishes persistence of the Hund dependence at lower cutoffs, not a zero-regulator limit or loop-order convergence. {The loop-order comparison in main-text Fig.~5(c,d) instead holds the regulator endpoint fixed.}

\section*{{Supplementary Note 8. Loop-order dependence of the altermagnetic results}}
\label{amsup:am-loop-order}

\par
Main-text Fig.~5(c,d) compares one to four loops at $R=10$, $T/U=10^{-2}$, $\Lambda\simeq0.5\Delta_{\mathrm{ref}}$, and $\chi=128$. Hund coupling reverses the self-energy asymmetry at every loop order. The zero obtained by linear interpolation between $J/U=0.10$ and $0.15$ lies at $J/U=0.105$, $0.110$, $0.109$, and $0.108$ at one to four loops, and the one-loop calculations at $R=7$ and $8$ change sign between the same two Hund couplings (Supplementary Note~5). The magnitude is not converged in loop order: two-loop corrections increase $|\overline{\delta\Sigma}|$, three- and four-loop corrections reduce it, and the four-loop value is smaller than the three-loop value by $8$--$14\%$; $|\operatorname{Im}\Sigma_\pm(i\pi T)|$ themselves decrease by $11$--$14\%$ for $J/U\ge0.15$ between three and four loops. The reversal therefore persists while its magnitude remains loop dependent.

The spin polarization of the orbital-field response increases with Hund coupling at every loop order [Fig.~5(d)]. At one loop it exceeds one at $J/U=0.30$, where the $xz\downarrow$ occupancy change reverses sign, so at strong Hund coupling the one-loop value lies far above the two- to four-loop values. Two to four loops agree to within $5\%$ for $J/U\le0.15$ and give $0.44$, $0.67$, and $0.55$ at $J/U=0.30$; the sequence alternates with decreasing amplitude but is not converged, so the magnitude of the polarization at strong Hund coupling remains loop dependent. All loop orders lie well above the dressed bubbles, which remain between $0.083$ and $0.095$ at all Hund couplings, so the strong Hund dependence requires vertex contributions and persists at four loops.

All comparisons use the same temperature, frequency range, regulator endpoint, and maximum bond dimensions, and each loop order is available at $J/U=0,0.10,0.15,0.20,0.25,0.30$. The endpoint bond dimension of the vertex state reaches $62$--$88$ at one loop and $88$--$118$ at two to four loops, below the prescribed $\chi=128$.
\par

\enlargethispage{\baselineskip}

\end{document}